\documentclass[11pt,oneside]{article}
\usepackage{amssymb}
\newcommand{\text}[1]{\mbox{#1}}
\newcommand{\binom}[2]{{#1 \choose #2}}
\newcommand{\QATOP}[2]{{#1 \atop #2}}
\newcommand{\QATOPD}[2]{\left[ \begin{array}{c} #1 \\ #2
    \end{array} \right]}
\begin{document}

\author{Hartmut Wachter\thanks{%
e-mail:Hartmut.Wachter@physik.uni-muenchen.de} \\
%EndAName
Sektion Physik, Ludwig-Maximilians-Universit\"{a}t,\\
Theresienstr. 37, D-80333 M\"{u}nchen, Germany}
\title{q-Integration on Quantum Spaces }
\maketitle

\begin{abstract}
In this article we present explicit formulae for q-integration on
quantum spaces which could be of particular importance in physics, i.e.,
q-deformed Minkowski space and q-deformed Euclidean space in three or four
dimensions. Furthermore, our formulae can be
regarded as a generalization of Jackson's q-integral to three and
four dimensions and provide a new possibility for an integration over
the whole space being invariant under translations and rotations.

\end{abstract}

\section{Introduction}

The history of the natural sciencies shows us that revolutionary changes in our
understanding of physical phenomena have often been accompanied by new
developments in mathematics. Newtonian mechanics is one famous example for
such a situation in physics, as its general formulation would not have been
able without the ideas of differential calculus. Due to this fact one may
believe that a new theory giving a more detailed description of nature has
to be based on a modified version of traditional mathematics.

Quantum spaces which are defined as co-module algebras of quantum groups and
which can be interpreted as deformations of ordinary co-ordinate algebras 
\cite{RTF90} could provide a proper framework for developing such a
modification \cite{Wes00}, \cite{Maj93}. For our \ purposes it is sufficient
to consider a quantum space as an algebra $\mathcal{A}_{q}$ of formal power
series in the non-commuting co-ordinates $X_{1},X_{2},\ldots ,X_{n}$

\begin{equation}
\mathcal{A}_{q}=\mathbb{C}\left[ \left[ X_{1},\ldots X_{n}\right] \right] /%
\mathcal{I}
\end{equation}
where $\mathcal{I}$ denotes the ideal generated by the relations of the
non-commuting co-ordinates.

The algebra $\mathcal{A}_{q}$ satisfies the Poincar\'{e}-Birkhoff-Witt
property, i.e. the dimension of the subspace of homogenous polynomials
should be the same as for commuting co-ordinates. This property is the
deeper reason why the monomials of normal ordering $X_{1}X_{2}\ldots X_{n}$
constitute a basis of $\mathcal{A}_{q}$. In particular, we can establish a
vector space isomorphism between $\mathcal{A}_{q}$ and the commutative
algebra $\mathcal{A}$ generated by ordinary co-ordinates $x_{1},x_{2},\ldots
,x_{n}$: 
\begin{eqnarray}
\mathcal{W} &:&\mathcal{A}\longrightarrow \mathcal{A}_{q}, \\
\mathcal{W}(x_{1}^{i_{1}}\ldots x_{n}^{i_{n}}) &=&X_{1}^{i_{1}}\ldots
X_{n}^{i_{n}}.  \nonumber
\end{eqnarray}
This vector space isomorphism can be extended to an algebra isomorphism
introducing a non-commutative product in $\mathcal{A}$, the so-called $\star 
$-product \cite{Moy49}, \cite{MSSW00}. This product is defined by the
relation 
\begin{equation}
\mathcal{W}(f\star g)=\mathcal{W}(f)\cdot \mathcal{W}(g)
\end{equation}
where $f$ and $g$ are formal power series in $\mathcal{A}$. In \cite{WW01}
we have calculated the $\star $-product for quantum spaces which could be of
particular importance in physics, i.e. q-deformed Minkowski space and
q-deformed Euclidean space in three or four dimensions.

Additionally, for each of these quantum spaces exists a symmetry algebra 
\cite{Dri85}, \cite{Jim85} and a covariant differential calculus \cite{WZ91}%
, which can provide an action upon the quantum spaces under consideration.
By means of the relation 
\begin{equation}
\mathcal{W}(h\triangleright f):=h\triangleright \mathcal{W}(f)\text{,\quad }%
h\in \mathcal{H}\text{, }f\in \mathcal{A}\text{,}
\end{equation}
we are also able to introduce an action upon the corresponding commutative
algebra. In our previous work \cite{BW01} we have presented explicit
formulae for such representations.

In the following it is our aim to derive some sort of q-integration on
commutative algebras which are, via the algebra isomorphism $\mathcal{W}$,
related to q-deformed Minkowski space or q-deformed Euclidean space in three
and four dimensions. In some sense our considerations can be supposed
to be a
generalization of the celebrated Jackson-integral \cite{Jac27} to higher
dimensions. To this end we will start with introducing new elements which
are formally inverse to the partial derivatives of the given covariant
differential calculi. Such an extension of the algebra of partial
derivatives will lead to additional commutation relations. Finally, the
representations of the partial derivatives given in \cite{BW01} will aid us
in identifying this new elements with particular solutions to some
q-difference equations. In this way, we can interpret our results as a
method to discretise classical integrals of more than one dimension.

Furthermore, we will see that we have to distinguish between left and right
integrals. For both types formulae for integration by parts can be derived.
It is also possible to define volume integrals which are invariant under
translations or the action of symmetry generators if surface terms are
neglected. Suprisingly, these integrals obey a rather simple structure, as
the single integrations for the different directions become independent from
each other if we integrate over the entire space.

\section{q-Deformed Euclidean space in three dimensions\label{DreiEukl}}

From \cite{LWW97} we know that the partial derivatives $\partial ^{+},$ $%
\partial ^{3},$ $\partial ^{-}$ satisfy the same relations as the quantum
space coordinates $X^{+},$ $X^{3},$ $X^{-}.$ Thus we have 
\begin{equation}
\partial ^{3}\partial ^{+}=q^{2}\partial ^{+}\partial ^{3},\quad \partial
^{-}\partial ^{3}=q^{2}\partial ^{3}\partial ^{-},\quad \partial
^{-}\partial ^{+}=\partial ^{+}\partial ^{-}+\lambda (\partial ^{3})^{2},
\end{equation}
where $\lambda =q-q^{-1}$ and $q>1$. Now, we would like to extend the
algebra of partial derivatives by inverse elements, such that 
\begin{eqnarray}
\partial ^{+}(\partial ^{+})^{-1} &=&(\partial ^{+})^{-1}\partial ^{+}=1, \\
\partial ^{3}(\partial ^{3})^{-1} &=&(\partial ^{3})^{-1}\partial ^{3}=1, 
\nonumber \\
\partial ^{-}(\partial ^{-})^{-1} &=&(\partial ^{-})^{-1}\partial ^{-}=1. 
\nonumber
\end{eqnarray}
It can easily be shown that these relations imply the formulae 
\begin{eqnarray}
(\partial ^{3})^{-1}\partial ^{+} &=&q^{-2}\partial ^{+}(\partial
^{3})^{-1},\quad \partial ^{3}(\partial ^{+})^{-1}=q^{-2}(\partial
^{+})^{-1}\partial ^{3}, \\
(\partial ^{-})^{-1}\partial ^{3} &=&q^{-2}\partial ^{3}(\partial
^{-})^{-1},\quad \partial ^{-}(\partial ^{3})^{-1}=q^{-2}(\partial
^{3})^{-1}\partial ^{-},  \nonumber \\
(\partial ^{-})^{-1}\partial ^{+} &=&\partial ^{+}(\partial
^{-})^{-1}-q^{-4}\lambda (\partial ^{3})^{2}(\partial ^{-})^{-2},  \nonumber
\\
\partial ^{-}(\partial ^{+})^{-1} &=&(\partial ^{+})^{-1}\partial
^{-}-q^{-4}\lambda (\partial ^{+})^{-2}(\partial ^{3})^{2}.  \nonumber
\end{eqnarray}
In addition, we can also find the following identities: 
\begin{eqnarray}
(\partial ^{3})^{-1}(\partial ^{+})^{-1} &=&q^{2}(\partial
^{+})^{-1}(\partial ^{3})^{-1},\\
 (\partial ^{-})^{-1}(\partial
^{3})^{-1}&=&q^{2}(\partial ^{3})^{-1}(\partial ^{-})^{-1}, \nonumber \\
(\partial ^{-})^{-1}(\partial ^{+})^{-1} &=&\sum_{i=0}^{\infty }\lambda
^{i}\left[ \left[ i\right] \right] _{q^{4}}!\left( \QATOPD {-1}{i}%
_{q^{4}} \right) ^{2} \nonumber \\
&& \cdot\, (\partial ^{+})^{-(i+1)}(\partial ^{3})^{2i}(\partial
^{-})^{-(i+1)}.  \nonumber
\end{eqnarray}
From the commutation relations between symmetry generators and partial
derivatives \cite{LWW97} we find the expressions 
\begin{eqnarray}
L^{+}(\partial ^{+})^{-1} &=&(\partial ^{+})^{-1}L^{+}, \\
L^{+}(\partial ^{3})^{-1} &=&(\partial ^{3})^{-1}L^{+}+q^{-1}\partial
^{+}(\partial ^{3})^{-2}\tau ^{-\frac{1}{2}},  \nonumber \\
L^{+}(\partial ^{-})^{-1} &=&(\partial ^{-})^{-1}L^{+}+q^{-1}\partial
^{3}(\partial ^{-})^{-2}\tau ^{-\frac{1}{2}},  \nonumber \\[0.1in]
L^{-}(\partial ^{-})^{-1} &=&(\partial ^{-})^{-1}L^{-}, \\
L^{-}(\partial ^{3})^{-1} &=&(\partial ^{3})^{-1}L^{-}-q^{-3}(\partial
^{3})^{-2}\partial ^{-}\tau ^{-\frac{1}{2}},  \nonumber \\
L^{-}(\partial ^{+})^{-1} &=&(\partial ^{+})^{-1}L^{-}-q^{-4}(\partial
^{+})^{-2}\partial ^{3}\tau ^{-\frac{1}{2}},  \nonumber \\[0.1in]
\tau ^{-\frac{1}{2}}(\partial ^{+})^{-1} &=&q^{-2}(\partial ^{+})^{-1}\tau
^{-\frac{1}{2}}, \\
\tau ^{-\frac{1}{2}}(\partial ^{-})^{-1} &=&q^{2}(\partial ^{-})^{-1}\tau ^{-%
\frac{1}{2}},  \nonumber \\
\tau ^{-\frac{1}{2}}(\partial ^{3})^{-1} &=&(\partial ^{3})^{-1}\tau ^{-%
\frac{1}{2}},  \nonumber \\[0.1in]
\Lambda ^{\frac{1}{2}}(\partial ^{A})^{-1} &=&q^{2}(\partial
^{A})^{-1}\Lambda ^{\frac{1}{2}},\quad A=\pm ,3.
\end{eqnarray}
Applying the substitutions 
\begin{equation}
\partial ^{A}\rightarrow \hat{\partial}^{A},\quad (\partial
^{A})^{-1}\rightarrow (\hat{\partial}^{A})^{-1},\quad A=\pm ,3
\end{equation}
to all expressions presented so far we get the corresponding relations of
the second differential calculus (generated by the conjugated partial
derivatives $\hat{\partial}^{A}$ \cite{LWW97}).

In \cite{BW01} it was shown that according to 
\begin{equation}
\partial ^{A}\triangleright F=\left( (\partial _{(i=0)}^{A})+(\partial
_{(i>0)}^{A})\right) F  \label{SplitAbl}
\end{equation}
the representations of our partial derivatives can be divided up into a
classical part and corrections vanishing in the undeformed limit $%
q\rightarrow 1$. Thus, seeking a solution to the equation 
\begin{equation}
\partial ^{A}\triangleright F=f  \label{DiffEuq}
\end{equation}
for given f it is reasonable to consider the following expression: 
\begin{eqnarray}
F &=&\left( \partial ^{A}\right) ^{-1}\triangleright f=\frac{1}{(\partial
_{(i=0)}^{A})+(\partial _{(i>0)}^{A})}f  \label{IntegralE3} \\
&=&\frac{1}{(\partial _{(i=0)}^{A})\left( 1+(\partial
_{(i=0)}^{A})^{-1}(\partial _{(i>0)}^{A})\right) }f  \nonumber \\
&=&\frac{1}{1+(\partial _{(i=0)}^{A})^{-1}(\partial _{(i>0)}^{A})}\cdot 
\frac{1}{(\partial _{(i=0)}^{A})}f  \nonumber \\
&=&\sum_{k=0}^{\infty }\left( -1\right) ^{k}\left[ (\partial
_{(i=0)}^{A})^{-1}(\partial _{(i>0)}^{A})\right] ^{k}(\partial
_{(i=0)}^{A})^{-1}f.  \nonumber
\end{eqnarray}
To apply this formula, we need to identify the contributions $\partial
_{(i=0)}^{A}$ and $\partial _{(i>0)}^{A}$ the representations of our partial
derivatives consist of. However, their explicit form can be read off quite
easily from the results in \cite{BW01}. Hence we have 
\begin{eqnarray}
(\partial _{\left( i=0\right) }^{+})f &=&-qD_{q^{4}}^{-}f(q^{2}x^{3}), \\
(\partial _{\left( i=0\right) }^{3})f &=&D_{q^{2}}^{3}f(q^{2}x^{+}), 
\nonumber \\
(\partial _{\left( i=0\right) }^{-})f &=&-q^{-1}D_{q^{4}}^{+}f  \nonumber
\end{eqnarray}
and 
\begin{eqnarray}
(\partial _{\left( i>0\right) }^{+})f &=&-q\lambda x^{+}(D_{q^{2}}^{3})^{2}f,
\\
(\partial _{\left( i>0\right) }^{3})f &=&0,  \nonumber \\
(\partial _{\left( i>0\right) }^{-})f &=&0.  \nonumber
\end{eqnarray}
Furthermore, it is easily seen that the inverse operators $(\partial
_{(i=0)}^{A})^{-1}$ are given by 
\begin{eqnarray}
(\partial _{\left( i=0\right) }^{+})^{-1}f
&=&-q^{-1}(D_{q^{4}}^{-})^{-1}f(q^{-2}x^{3}), \\
(\partial _{\left( i=0\right) }^{3})^{-1}f
&=&(D_{q^{2}}^{3})^{-1}f(q^{-2}x^{+}),  \nonumber \\
(\partial _{\left( i=0\right) }^{-})^{-1}f &=&-q(D_{q^{4}}^{+})^{-1}f, 
\nonumber
\end{eqnarray}
where $D_{q^{a}}^{A}$ and $(D_{q^{a}}^{A})^{-1}$ denote Jackson derivatives
and Jackson integrals, respectively. Inserting these expressions into
formula (\ref{IntegralE3}) we can represent the inverse counterparts of the
partial derivatives in the form

\begin{eqnarray}
(\partial ^{-})_{L}^{-1}f &=&-q(D_{q^{4}}^{+})^{-1}f,  \label{repres1} \\
(\partial ^{3})_{L}^{-1}f &=&(D_{q^{2}}^{3})^{-1}f(q^{-2}x^{+}),  \nonumber
\\
(\partial ^{+})_{L}^{-1}f &=&-q^{-1}\sum_{k=0}^{\infty }(-\lambda
)^{k}q^{2k(k+1)}\left[ x^{+}(D_{q^{4}}^{-})^{-1}(D_{q^{2}}^{3})^{2}\right]
^{k}  \nonumber \\
&&\cdot {}\,(D_{q^{4}}^{-})^{-1}f(q^{-2(k+1)}x^{3}).  \nonumber
\end{eqnarray}
Repeating the identical steps as before, we can also find solutions to the
equations 
\begin{eqnarray}
\hat{\partial}^{A}\,\tilde{\triangleright}\,F &=&f,  \label{DiffEuqCon} \\
F\,\tilde{\triangleleft}\,\partial ^{A} &=&f,  \nonumber \\
F\triangleleft \hat{\partial}^{A} &=&f,  \nonumber
\end{eqnarray}
where the explicit form of the action of our partial derivatives is again
taken from \cite{BW01}. However, in \cite{BW01} it was also shown that the
different representations of our partial derivatives can be transformed into
each other by a straightforward application of so-called
crossing-symmetries. Due to this fact the solutions to equation (\ref
{DiffEuqCon}) are easily obtained from (\ref{repres1}), if we apply the
transformations 
\begin{eqnarray}
(\partial ^{\pm })^{-1}\triangleright f&\stackrel{{\QATOP{\pm }{q}}{\QATOP{%
\rightarrow }{\rightarrow }}{\QATOP{\mp }{1/q}}}{\longleftrightarrow }&(\hat{%
\partial}^{\mp })^{-1}\tilde{\triangleright}f,  \label{TrafoUnKon} \\
(\partial ^{3})^{-1}\triangleright f&\stackrel{{\QATOP{\pm }{q}}{\QATOP{%
\rightarrow }{\rightarrow }}{\QATOP{\mp }{1/q}}}{\longleftrightarrow }&(\hat{%
\partial}^{3})^{-1}\tilde{\triangleright}f,  \nonumber
\end{eqnarray}
which concretely mean, that the expressions on the right \ and left hand
side are related to each other by the substitutions\footnote{%
For notation see appendix \ref{AppA}, please.} 
\begin{equation}
x^{\pm }\rightarrow x^{\mp },\quad q^{\pm 1}\rightarrow q^{\mp 1},\quad \hat{%
n}^{\pm }\rightarrow -\hat{n}^{\mp }  \label{TrafoUnKonexpl}
\end{equation}
\[
D_{q^{a}}^{\pm }\rightarrow D_{q^{-a}}^{\mp },\quad (D_{q^{a}}^{\pm
})^{-1}\rightarrow (D_{q^{-a}}^{\mp })^{-1}. 
\]
In the same way we have 
\begin{eqnarray}
f\,\tilde{\triangleleft}\,(\partial ^{\pm })^{-1}
&\stackrel{+\leftrightarrow -}{%
\longleftrightarrow }&-q^{6}(\hat{\partial}^{\mp })^{-1}
\,\tilde{\triangleright}\,%
f,  \label{TrafoLinRech1} \\
f\,\tilde{\triangleleft}\,(\partial ^{3})^{-1}&\stackrel{+\leftrightarrow -}{%
\longleftrightarrow }&-q^{6}(\hat{\partial}^{3})^{-1}
\,\tilde{\triangleright}\,f,
\nonumber \\
f\triangleleft (\hat{\partial}^{\pm })^{-1}&\stackrel{+\leftrightarrow -}{%
\longleftrightarrow }&-q^{-6}(\partial ^{\mp })^{-1}\triangleright f,
\label{TrafoLinRech2} \\
f\triangleleft (\hat{\partial}^{3})^{-1}&\stackrel{+\leftrightarrow -}{%
\longleftrightarrow }&-q^{-6}(\partial ^{3})^{-1}\triangleright f,  \nonumber
\end{eqnarray}
which symbolizes a transition described by the substitutions 
\begin{equation}
x^{\pm }\rightarrow x^{\mp },\quad D_{q^{a}}^{\pm }\rightarrow
D_{q^{a}}^{\mp },\quad (D_{q^{a}}^{\pm })^{-1}\rightarrow (D_{q^{a}}^{\mp
})^{-1},\quad \hat{n}^{\pm }\rightarrow \hat{n}^{\mp }.
\end{equation}

Next, we would like to have a closer look at the question in which sense
the operator $(\partial ^{A})^{-1}$ inverse to the operator $\partial
^{A}$ is.
First of all, we require that all lower limits of Jackson integrals
appearing in the representations of $(\partial ^{A})^{-1}$ are set equal to
zero. In this case one can easily verify the identities 
\begin{eqnarray}
D_{q^{a}}^{A}\left[ \left. (D_{q^{a}}^{A})^{-1}f\right| _{0}^{x^{A}}\right] 
&=&f, \\
\left. (D_{q^{a}}^{A})^{-1}D_{q^{a}}^{A}f\right| _{0}^{x^{A}} &=&\left.
f\right| _{0}^{x^{A}}  \nonumber
\end{eqnarray}
leading to 
\begin{eqnarray}
\partial _{\left( i=0\right) }^{A}\left[ \left. (\partial _{\left(
i=0\right) }^{A})^{-1}f\right| _{0}^{x^{\bar{A}}}\right]  &=&f,
\label{HauDifInt} \\
\left. (\partial _{\left( i=0\right) }^{A})^{-1}\partial _{\left( i=0\right)
}^{A}f\right| _{0}^{x^{\bar{A}}} &=&\left. f\right| _{0}^{x^{\bar{A}}}. 
\nonumber
\end{eqnarray}
It is important to realize, that in the above formulae the coordinates
giving the upper limits of integration have to be labeled by different
indices than the corresponding operators $(\partial _{\left( i=0\right)
}^{A})^{-1}.$ Let us now introduce the operator 
\begin{equation}
\hat{C}^{A}f=-\left. (\partial _{\left( i=0\right) }^{A})^{-1}(\partial
_{\left( i>0\right) }^{A})f\right| _{0}^{x^{\bar{A}}}.
\end{equation}
With the identities (\ref{SplitAbl}), (\ref{IntegralE3}) and (\ref{HauDifInt}%
) one can then compute that 
\begin{eqnarray}
&&\left. (\partial ^{A})^{-1}\partial ^{A}f\right| _{0}^{x^{\bar{A}}}
\label{NichHauDifInt} \\
&=&\sum_{k=0}^{\infty }\left. (\hat{C}^{A})^{k}(f-f(x^{\bar{A}}=0)\right|
_{0}^{x^{\bar{A}}}-\sum_{k=1}^{\infty }\left. (\hat{C}^{A})^{k}f\right|
_{0}^{x^{\bar{A}}}  \nonumber \\
&=&f-f(x^{\bar{A}}=0)-\sum_{k=1}^{\infty }\left. (\hat{C}^{A})^{k}f(x^{\bar{A%
}}=0)\right| _{0}^{x^{\bar{A}}}  \nonumber
\end{eqnarray}
where the limits of the integration intervals \emph{always refer to the
Jackson integrals} in the expressions of $(\partial ^{A})^{-1}$ or $\hat{C}%
^{A}$. In addition to this one readily checks that 
\begin{equation}
\partial ^{A}\left[ \left. (\partial ^{A})^{-1}f\right| _{0}^{x^{\bar{A}%
}}\right] =f,  \label{2VerHauptDifInt}
\end{equation}
as the representations of $(\partial ^{A})^{-1}$ have been determined in
such a way that they give solutions to equation (\ref{DiffEuq}).

As a next step we wish to present formulae for integration by parts. Before
doing this let us collect some notation that will be used in the following.
In the case of right and left derivatives we use the abbreviations 
\begin{eqnarray}
\partial _{L}^{A}f &\equiv &\partial ^{A}\triangleright f, \\
\partial _{R}^{A}f &\equiv &f\,\tilde{\triangleleft}\,\partial ^{A},  \nonumber
\\[0.16in]
\hat{\partial}_{L}^{A}f &\equiv &\hat{\partial}^{A}\,\tilde{\triangleright}\,f,
\\
\hat{\partial}_{R}^{A}f &\equiv &f\triangleleft \hat{\partial}^{A}. 
\nonumber
\end{eqnarray}
Likewise for right and left integrals we abbreviate 
\begin{eqnarray}
(\partial ^{A})_{L}^{-1}f &\equiv &(\partial ^{A})^{-1}\triangleright f, \\
(\partial ^{A})_{L}^{-1}f &\equiv &f\,\tilde{\triangleleft}\,(\partial
^{A})^{-1},  \nonumber \\[0.16in]
(\hat{\partial}^{A})_{R}^{-1}f &\equiv &(\hat{\partial}^{A})^{-1}\,\tilde{%
\triangleright}\,f, \\
(\hat{\partial}^{A})_{R}^{-1}f &\equiv &f\triangleleft (\hat{\partial}%
^{A})^{-1}.  \nonumber
\end{eqnarray}
Furthermore, we set 
\begin{equation}
\left. f\right\| _{0}^{x^{\bar{A}}}\equiv f-f(x^{\bar{A}}=0)-\sum_{k=1}^{%
\infty }(\hat{C}^{A})^{k}\left. f(x^{\bar{A}}=0)\right| _{0}^{x^{\bar{A}}}.
\end{equation}
Now, we are in a position to write down rules for integration by parts. In
complete analogy to the classical case these rules can be derived from the
Leibniz rules of the corresponding partial derivatives \cite{BW01}. In this
way we get 
\begin{eqnarray}
&&(\partial ^{-})_{L}^{-1}(\partial _{L}^{-}f)\star g\left|
_{x^{+}=0}^{a}\right.\\&=&f\star g\left\| _{x^{+}=0}^{a}\right.  
-{}(\partial ^{-})_{L}^{-1}(\Lambda ^{1/2}\tau ^{-1/2}f)\star \partial
_{L}^{-}g\left| _{x^{+}=0}^{a},\right.  \nonumber  \\[0.16in]
&&(\partial ^{3})_{L}^{-1}(\partial _{L}^{3}f)\star g\left|
_{x^{3}=0}^{a}\right.\\&=&f\star g\left\| _{x^{3}=0}^{a}\right. 
-(\partial
^{3})_{L}^{-1}(\Lambda ^{1/2}f)\star \partial _{L}^{3}g\left|
_{x^{3}=0}^{a}\right.  \nonumber \\
&&-\,\lambda \lambda _{+}(\partial ^{3})_{L}^{-1}(\Lambda ^{1/2}L^{+}f)\star
\partial _{L}^{-}g\left| _{x^{3}=0}^{a}\right. ,  \nonumber \\[0.16in]
&&(\partial ^{+})_{L}^{-1}(\partial _{L}^{+}f)\star g\left|
_{x^{-}=0}^{a}\right.\\ &=&f\star g\left\| _{x^{-}=0}^{a}\right.
-(\partial
^{+})_{L}^{-1}(\Lambda ^{1/2}\tau ^{1/2}f)\star \partial _{L}^{+}g\left|
_{x^{-}=0}^{a}\right.  \nonumber \\
&&-\,q\lambda \lambda _{+}(\partial ^{+})_{L}^{-1}(\Lambda ^{1/2}\tau
^{1/2}L^{+}f)\star \partial _{L}^{3}g\left| _{x^{-}=a}^{b}\right.  \nonumber
\\
&&-\,q^{2}\lambda ^{2}\lambda _{+}(\partial ^{+})_{L}^{-1}(\Lambda
^{1/2}\tau ^{1/2}\left( L^{+}\right) ^{2}f)\star \partial _{L}^{-}g\left|
_{x^{-}=a}^{b}\right.  \nonumber
\end{eqnarray}
and 
\begin{eqnarray}
&&(\hat{\partial}^{+})_{L}^{-1}(\hat{\partial}_{L}^{+}f)\star g\left|
_{x^{-}=0}^{a}\right.\\ &=&f\star g\left\| _{x^{-}=0}^{a}\right. 
-{}(\hat{\partial}^{+})_{L}^{-1}(\Lambda ^{-1/2}\tau ^{-1/2}f)\star \hat{%
\partial}_{L}^{+}g\left| _{x^{-}=0}^{a},\right.  \nonumber \\[0.16in]
&&(\hat{\partial}^{3})_{L}^{-1}(\hat{\partial}_{L}^{3}f)\star g\left|
_{x^{3}=0}^{a}\right.\\ &=&f\star g\left\| _{x^{3}=0}^{a}\right.-(\hat{%
\partial}^{3})_{L}^{-1}(\Lambda ^{-1/2}f)\star \hat{\partial}_{L}^{3}g\left|
_{x^{3}=0}^{a}\right.  \nonumber \\
&&-\,\lambda \lambda _{+}(\hat{\partial}^{3})_{L}^{-1}\left( \Lambda
^{-1/2}L^{-}f\right) \star \hat{\partial}_{L}^{+}g\left|
_{x^{3}=0}^{a}\right. ,  \nonumber \\[0.16in]
&&(\hat{\partial}^{-})_{L}^{-1}(\hat{\partial}_{L}^{-}f)\star g\left|
_{x^{+}=0}^{a}\right.\\ &=&f\star g\left\| _{x^{+}=0}^{a}\right.
-(\hat{%
\partial}^{-})_{L}^{-1}(\Lambda ^{-1/2}\tau ^{1/2}f)\star \hat{\partial}%
_{L}^{-}g\left| _{x^{+}=0}^{a}\right.  \nonumber \\
&&-\,q^{-1}\lambda \lambda _{+}(\hat{\partial}^{-})_{L}^{-1}(\Lambda
^{-1/2}\tau ^{1/2}L^{-}f)\star \hat{\partial}_{L}^{3}g\left|
_{x^{+}=0}^{a}\right.  \nonumber \\
&&-\,q^{-2}\lambda ^{2}\lambda _{+}(\hat{\partial}^{-})_{L}^{-1}(\Lambda
^{-1/2}\tau ^{1/2}(L^{-})^{2}f)\star \hat{\partial}_{L}^{+}g\left|
_{x^{+}=0}^{a}\right. .  \nonumber
\end{eqnarray}
In the case of right integrals, however, we have to consider Leibniz rules
for right derivatives \cite{BW01}. Hence, we have 
\begin{eqnarray}
&&(\partial ^{-})_{R}^{-1}f\star \partial _{R}^{-}g\left|
_{x^{+}=0}^{a}\right.\\ &=&f\star g\left\| _{x^{+}=0}^{a}\right. 
-{}(\partial ^{-})_{R}^{-1}(\partial _{R}^{-}f)\star (\Lambda ^{-1/2}\tau
^{1/2}g)\left| _{x^{+}=0}^{a},\right.  \nonumber \\[0.16in]
&&(\partial ^{3})_{R}^{-1}f\star \partial _{R}^{3}g\left|
_{x^{3}=0}^{a}\right.\\&=&f\star g\left\| _{x^{3}=0}^{a}\right. -(\partial
^{3})_{R}^{-1}(\partial _{R}^{3}f)\star (\Lambda ^{-1/2}g)\left|
_{x^{3}=0}^{a}\right.  \nonumber \\
&&+\,\lambda \lambda _{+}(\partial ^{3})_{R}^{-1}(\partial _{R}^{-}f)\star
(\Lambda ^{-1/2}\tau ^{1/2}L^{+}g)\left| _{x^{3}=0}^{a},\right.
\nonumber
 \\[0.16in]
&&(\partial ^{+})_{R}^{-1}f\star \partial _{R}^{+}g\left|
_{x^{-}=0}^{a}\right.\\ &=&f\star g\left\| _{x^{-}=0}^{a}\right.
-(\partial
^{+})_{R}^{-1}(\partial _{R}^{+}f)\star (\Lambda ^{-1/2}\tau ^{-1/2}g)\left|
_{x^{-}=0}^{a}\right.  \nonumber \\
&&+\,q^{-1}\lambda \lambda _{+}(\partial ^{+})_{R}^{-1}(\partial
_{R}^{3}f)\star (\Lambda ^{-1/2}L^{+}g)\left| _{x^{-}=0}^{a}\right. 
\nonumber \\
&&-\,\lambda ^{2}\lambda _{+}(\partial ^{+})_{R}^{-1}(\partial
_{R}^{-}f)\star (\Lambda ^{-1/2}\tau ^{1/2}\left( L^{+}\right) ^{2}g)\left|
_{x^{-}=0}^{a}\right.  \nonumber
\end{eqnarray}
and 
\begin{eqnarray}
&&(\hat{\partial}^{+})_{R}^{-1}f\star \hat{\partial}_{R}^{+}g\left|
_{x^{-}=0}^{a}\right.\\ &=&f\star g\left\| _{x^{-}=0}^{a}\right. 
-{}(\hat{\partial}^{+})_{R}^{-1}(\hat{\partial}_{R}^{+}f)\star (\Lambda
^{1/2}\tau ^{1/2}g)\left| _{x^{-}=0}^{a},\right.  \nonumber \\[0.16in]
&&(\hat{\partial}^{3})_{R}^{-1}f\star \hat{\partial}_{R}^{3}g\left|
_{x^{3}=0}^{a}\right.\\ &=&f\star g\left\| _{x^{3}=0}^{a}\right.
-(\hat{%
\partial}^{3})_{R}^{-1}(\hat{\partial}_{R}^{3}f)\star (\Lambda
^{1/2}g)\left| _{x^{3}=0}^{a}\right.  \nonumber \\
&&+\,\lambda \lambda _{+}(\hat{\partial}^{3})_{R}^{-1}(\hat{\partial}%
_{R}^{+}f)\star (\Lambda ^{1/2}\tau ^{1/2}L^{-}g)\left|
_{x^{3}=0}^{a},\right.  \nonumber \\[0.16in]
&&(\hat{\partial}^{-})_{R}^{-1}f\star \hat{\partial}_{R}^{-}g\left|
_{x^{+}=0}^{a}\right.\\ &=&f\star g\left\| _{x^{+}=0}^{a}\right.
-(\hat{%
\partial}^{-})_{R}^{-1}(\hat{\partial}_{R}^{-}f)\star (\Lambda ^{1/2}\tau
^{-1/2}g)\left| _{x^{+}=0}^{a}\right.  \nonumber \\
&&+\,q\lambda \lambda _{+}(\hat{\partial}^{-})_{R}^{-1}(\hat{\partial}%
_{R}^{3}f)\star (\Lambda ^{1/2}L^{-}g)\left| _{x^{+}=0}^{a}\right.  \nonumber
\\
&&-\,\lambda ^{2}\lambda _{+}(\hat{\partial}^{-})_{R}^{-1}(\hat{\partial}%
_{R}^{+}f)\star (\Lambda ^{1/2}\tau ^{1/2}\left( L^{-}\right) ^{2}g)\left|
_{x^{+}=0}^{a}\right. .  \nonumber
\end{eqnarray}

We want to close this section by discussing a method for an integration over
the entire space. For this purpose we could define 
\begin{equation}
\int_{L}d_{q}V\,f\equiv (\partial ^{+})^{-1}(\partial ^{3})^{-1}(\partial
^{-})^{-1}\triangleright f
\end{equation}
for a 3-dimensional volume integral. But this is not the only possibility
for defining a global integration, as the operators $(\partial ^{A})^{-1},$ $%
A=\pm ,3,$ do not mutually commute. However, for functions vanishing at
infinity one can show by inserting the representations of (\ref{repres1})
that the following indentities hold: 
\begin{eqnarray}
&&(\partial ^{+})^{-1}(\partial ^{3})^{-1}(\partial ^{-})^{-1}\triangleright
f \\
&=&q^{-4}(\partial ^{-})^{-1}(\partial ^{3})^{-1}(\partial
^{+})^{-1}\triangleright f+S.T.  \nonumber \\
&=&q^{-2}(\partial ^{3})^{-1}(\partial ^{-})^{-1}(\partial
^{+})^{-1}\triangleright f+S.T.  \nonumber \\
&=&q^{-2}(\partial ^{3})^{-1}(\partial ^{+})^{-1}(\partial
^{-})^{-1}\triangleright f+S.T.  \nonumber \\
&=&q^{-2}(\partial ^{-})^{-1}(\partial ^{+})^{-1}(\partial
^{3})^{-1}\triangleright f+S.T.  \nonumber \\
&=&q^{-2}(\partial ^{+})^{-1}(\partial ^{-})^{-1}(\partial
^{3})^{-1}\triangleright f+S.T.  \nonumber
\end{eqnarray}
where $S.T.$ stands for neglected surface terms. And in the same manner we
can also find as an explicit formula for calculating volume integrals 
\begin{eqnarray}
&&(\partial ^{+})^{-1}(\partial ^{3})^{-1}(\partial ^{-})^{-1}\triangleright
f  \label{DarVol3dim} \\
&=&(\partial _{(i=0)}^{+})^{-1}(\partial _{(i=0)}^{3})^{-1}(\partial
_{(i=0)}^{-})^{-1}\triangleright f+O.T.  \nonumber \\
&=&q^{-4}(D_{q^{4}}^{-})^{-1}(D_{q^{2}}^{3})^{-1}(D_{q^{4}}^{+})^{-1}f(q^{-2}x^{+},q^{-2}x^{3})+S.T.
\nonumber
\end{eqnarray}
if again surface terms are dropped. Using the crossing-symmetries (\ref
{TrafoUnKon}), (\ref{TrafoLinRech1}) and (\ref{TrafoLinRech2}) it
immediately follows that 
\begin{equation}
(\partial ^{+})^{-1}(\partial ^{3})^{-1}(\partial ^{-})^{-1}\triangleright f%
\stackrel{{\QATOP{\pm }{q}}{\QATOP{\rightarrow }{\rightarrow }}{\QATOP{\mp }{%
1/q}}}{\longleftrightarrow }(\hat{\partial}^{-})^{-1}(\hat{\partial}%
^{3})^{-1}(\hat{\partial}^{+})^{-1}\,\tilde{\triangleright}\,f
\end{equation}
and 
\begin{eqnarray}
f\,\tilde{\triangleleft}\,(\partial ^{+})^{-1}(\partial ^{3})^{-1}(\partial
^{-})^{-1}&\stackrel{+\leftrightarrow -}{\longleftrightarrow }&(\bar{\partial}%
^{-})^{-1}(\bar{\partial}^{3})^{-1}(\bar{\partial}^{+})^{-1}\,\tilde{%
\triangleright}\,f, \\
f\triangleleft (\bar{\partial}^{-})^{-1}(\bar{\partial}^{3})^{-1}(\bar{%
\partial}^{+})^{-1}&\stackrel{+\leftrightarrow -}{\longleftrightarrow }&%
(\partial ^{+})^{-1}(\partial ^{3})^{-1}(\partial ^{-})^{-1}\triangleright f,
\nonumber
\end{eqnarray}
where $(\bar{\partial}^{A})^{-1}=-q^{6}(\hat{\partial}^{A})^{-1}.$

From a direct calculation using the representations in \cite{BW01} and
appying them to the last expressions of (\ref{DarVol3dim}) one can also
verify that our volume integrals are invariant under both translations and
rotations. This means, in explicit form, that we have 
\begin{eqnarray}
\partial ^{A}\triangleright \int_{L}d_{q}V{}f &=&\int_{L}d_{q}V{}\partial
^{A}\triangleright f=\varepsilon (\partial ^{A})\int_{L}d_{q}V{}f=0,
\label{LinksInInt} \\
L^{\pm }\triangleright \int_{L}d_{q}V{}f &=&\int_{L}d_{q}V{}L^{\pm
}\triangleright f=\varepsilon (L^{\pm })\int_{L}d_{q}V{}f=0,  \nonumber \\
\tau \triangleright \int_{L}d_{q}V{}f &=&\int_{L}d_{q}V{}\tau \triangleright
f=\varepsilon (\tau )\int_{L}d_{q}V{}f=\int_{L}d_{q}V{}f,  \nonumber
\end{eqnarray}
where we have used as a shorthand notation 
\begin{eqnarray}
\int_{L}d_{q}V{}f &\equiv &(\partial ^{+})^{-1}(\partial ^{3})^{-1}(\partial
^{-})^{-1}\triangleright f, \\
\int_{R}d_{q}V{}f &\equiv &f\,\tilde{\triangleleft}\,(\partial
^{+})^{-1}(\partial ^{3})^{-1}(\partial ^{-})^{-1},  \nonumber \\
\int_{L}d_{q}\overline{V}{}f &\equiv &(\bar{\partial}^{-})^{-1}(\bar{\partial%
}^{3})^{-1}(\bar{\partial}^{+})^{-1}\,\tilde{\triangleright}\,f,  \nonumber \\
\int_{R}d_{q}\overline{V}{}f &\equiv &f\triangleleft (\bar{\partial}%
^{-})^{-1}(\bar{\partial}^{3})^{-1}(\bar{\partial}^{+})^{-1}.  \nonumber
\end{eqnarray}
In the same way one can check invariance under right representations,
hence 
\begin{eqnarray}
\left( \int_{L}d_{q}V{}f\right)\, \tilde{\triangleleft}\,\partial ^{A}
&=&\int_{L}d_{q}V{}f\,\tilde{\triangleleft}\,\partial ^{A}=\varepsilon (\partial
^{A})\int_{L}d_{q}V{}f=0, \\
\left( \int_{L}d_{q}V{}f\right) \triangleleft L^{\pm }
&=&\int_{L}d_{q}V{}f\triangleleft L^{\pm }=\varepsilon (L^{\pm
})\int_{L}d_{q}V{}f=0,  \nonumber \\
\left( \int_{L}d_{q}V{}f\right) \triangleleft \tau
&=&\int_{L}d_{q}V{}f\triangleleft \tau =\varepsilon (\tau
)\int_{L}d_{q}V{}f=\int_{L}d_{q}V{}f.  \nonumber
\end{eqnarray}
Because of the crossing-symmetries these indentities carry over to all of
the other volume integrals. However, one has to notice that in the case of
conjugated integrals translation invariance now reads 
\begin{eqnarray}
\bar{\partial}^{A}\triangleright \int_{L}d_{q}\overline{V}{}f
&=&\int_{L}d_{q}\overline{V}{}\bar{\partial}^{A}\triangleright f=\varepsilon
(\bar{\partial}^{A})\int_{L}d_{q}\overline{V}{}f=0, \\
\left( \int_{L}d_{q}\overline{V}{}f\right) \triangleleft \bar{\partial}^{A}
&=&\int_{L}d_{q}\overline{V}{}f\triangleleft \bar{\partial}^{A}=\varepsilon (%
\bar{\partial}^{A})\int_{L}d_{q}\overline{V}{}f=0,  \nonumber \\
\bar{\partial}^{A}\,\tilde{\triangleright}\int_{R}d_{q}\overline{V}{}f
&=&\int_{R}d_{q}\overline{V}{}\bar{\partial}^{A}\,\tilde{\triangleright}\,%
f=\varepsilon (\bar{\partial}^{A})\int_{R}d_{q}\overline{V}{}f=0,  \nonumber
\\
\left( \int_{R}d_{q}\overline{V}{}f\right) \triangleleft \bar{\partial}^{A}
&=&\int_{R}d_{q}\overline{V}{}f\triangleleft \bar{\partial}^{A}=\varepsilon (%
\bar{\partial}^{A})\int_{R}d_{q}\overline{V}{}f=0.  \nonumber
\end{eqnarray}

\section{q-Deformed Euclidean space in four dimensions}

The 4-dimensional Euclidean space can be treated in very much the same way
as the 3-dimensional one. Therefore we will restrict ourselves to stating the
results only. Again, we start with the commutation relations \cite{Oca96} 
\begin{equation}
\partial ^{1}\partial ^{2}=q\partial ^{2}\partial ^{1},\quad \partial
^{1}\partial ^{3}=q\partial ^{3}\partial ^{1},\quad \partial ^{2}\partial
^{4}=q\partial ^{4}\partial ^{2},\quad \partial ^{3}\partial ^{4}=q\partial
^{3}\partial ^{2},
\end{equation}
\[
\partial ^{2}\partial ^{3}=\partial ^{3}\partial ^{2},\quad \partial
^{4}\partial ^{1}=\partial ^{1}\partial ^{4}+\lambda \partial ^{2}\partial
^{3}, 
\]
where $\lambda =q-q^{-1}$ with $q>1$, and introduce elements $\left(
\partial ^{i}\right) ^{-1}$, $i=1,\ldots ,4,$ by 
\[
\partial ^{i}\left( \partial ^{i}\right) ^{-1}=\left( \partial ^{i}\right)
^{-1}\partial ^{i}=1,\quad i=1,\ldots ,4. 
\]
Now, the additional commutation relations become 
\begin{eqnarray}
(\partial ^{2})^{-1}\partial ^{1} &=&q\partial ^{1}(\partial
^{2})^{-1},\quad \partial ^{2}(\partial ^{1})^{-1}=q(\partial
^{1})^{-1}\partial ^{2}, \\
(\partial ^{3})^{-1}\partial ^{1} &=&q\partial ^{1}(\partial
^{3})^{-1},\quad \partial ^{3}(\partial ^{1})^{-1}=q(\partial
^{1})^{-1}\partial ^{3},  \nonumber \\
(\partial ^{4})^{-1}\partial ^{2} &=&q\partial ^{2}(\partial
^{4})^{-1},\quad \partial ^{4}(\partial ^{2})^{-1}=q(\partial
^{2})^{-1}\partial ^{4},  \nonumber \\
(\partial ^{4})^{-1}\partial ^{3} &=&q\partial ^{3}(\partial
^{4})^{-1},\quad \partial ^{4}(\partial ^{3})^{-1}=q(\partial
^{3})^{-1}\partial ^{4},  \nonumber \\
(\partial ^{4})^{-1}\partial ^{1} &=&\partial ^{1}(\partial
^{4})^{-1}-q^{2}\lambda \partial ^{2}\partial ^{3}(\partial ^{4})^{-2}, 
\nonumber \\
\partial ^{4}(\partial ^{1})^{-1} &=&(\partial ^{1})^{-1}\partial
^{4}-q^{2}\lambda (\partial ^{1})^{-2}\partial ^{2}\partial ^{3},  \nonumber
\\
(\partial ^{3})^{-1}\partial ^{2} &=&\partial ^{2}(\partial ^{3})^{-1},\quad
\partial ^{3}(\partial ^{2})^{-1}=(\partial ^{2})^{-1}\partial ^{3}. 
\nonumber
\end{eqnarray}
Furthermore, we obtain identities of the form 
\begin{eqnarray}
(\partial ^{2})^{-1}(\partial ^{1})^{-1} &=&q^{-1}(\partial
^{1})^{-1}(\partial ^{2})^{-1}, \\
(\partial ^{3})^{-1}(\partial ^{1})^{-1} &=&q^{-1}(\partial
^{1})^{-1}(\partial ^{3})^{-1},  \nonumber \\
(\partial ^{4})^{-1}(\partial ^{2})^{-1} &=&q^{-1}(\partial
^{2})^{-1}(\partial ^{4})^{-1},  \nonumber \\
(\partial ^{4})^{-1}(\partial ^{3})^{-1} &=&q^{-1}(\partial
^{3})^{-1}(\partial ^{4})^{-1},  \nonumber \\
(\partial ^{3})^{-1}(\partial ^{2})^{-1} &=&(\partial ^{2})^{-1}(\partial
^{3})^{-1},  \nonumber \\
(\partial ^{4})^{-1}(\partial ^{1})^{-1} &=&\sum_{i=0}^{\infty }\lambda
^{i}\left[ \left[ i\right] \right] _{q^{-2}}!\left( \QATOPD {-1}{i}%
_{q^{-2}}\right) ^{2}  \nonumber \\
&&\cdot\, {}(\partial ^{1})^{-(i+1)}(\partial ^{2})^{i}(\partial
^{3})^{i}(\partial ^{4})^{-(i+1)}.  \nonumber
\end{eqnarray}
Next, the commutation relations with the symmetry generators read 
\begin{eqnarray}
L_{1}^{+}\left( \partial ^{1}\right) ^{-1} &=&q^{-1}\left( \partial
^{1}\right) ^{-1}L_{1}^{+}+q^{-1}\left( \partial ^{1}\right) ^{-2}\partial
^{2}, \\
L_{1}^{+}\left( \partial ^{2}\right) ^{-1} &=&q\left( \partial ^{2}\right)
^{-1}L_{1}^{+},  \nonumber \\
L_{1}^{+}\left( \partial ^{3}\right) ^{-1} &=&q^{-1}\left( \partial
^{3}\right) ^{-1}L_{1}^{+}-q^{-1}\left( \partial ^{3}\right) ^{-2}\partial
^{4},  \nonumber \\
L_{1}^{+}\left( \partial ^{4}\right) ^{-1} &=&q\left( \partial ^{4}\right)
^{-1}L_{1}^{+},  \nonumber \\[0.16in]
L_{2}^{+}\left( \partial ^{1}\right) ^{-1} &=&q^{-1}\left( \partial
^{1}\right) ^{-1}L_{2}^{+}+q^{-1}\left( \partial ^{1}\right) ^{-2}\partial
^{3}, \\
L_{2}^{+}\left( \partial ^{2}\right) ^{-1} &=&q^{-1}\left( \partial
^{2}\right) ^{-1}L_{2}^{+}-q^{-1}\left( \partial ^{2}\right) ^{-2}\partial
^{4},  \nonumber \\
L_{2}^{+}\left( \partial ^{3}\right) ^{-1} &=&q\left( \partial ^{3}\right)
^{-1}L_{2}^{+},  \nonumber \\
L_{2}^{+}\left( \partial ^{4}\right) ^{-1} &=&q\left( \partial ^{4}\right)
^{-1}L_{2}^{+},  \nonumber \\[0.4cm]
L_{1}^{-}\left( \partial ^{1}\right) ^{-1} &=&q^{-1}\left( \partial
^{1}\right) ^{-1}L_{1}^{-}, \\
L_{1}^{-}\left( \partial ^{2}\right) ^{-1} &=&q\left( \partial ^{2}\right)
^{-1}L_{1}^{-}+q^{3}\partial ^{1}\left( \partial ^{2}\right) ^{-2}, 
\nonumber \\
L_{1}^{-}\left( \partial ^{3}\right) ^{-1} &=&q^{-1}\left( \partial
^{3}\right) ^{-1}L_{1}^{-},  \nonumber \\
L_{1}^{-}\left( \partial ^{4}\right) ^{-1} &=&q\left( \partial ^{4}\right)
^{-1}L_{1}^{-}-q^{3}\partial ^{3}\left( \partial ^{4}\right) ^{-2}, 
\nonumber \\[0.4cm]
L_{2}^{-}\left( \partial ^{1}\right) ^{-1} &=&q^{-1}\left( \partial
^{1}\right) ^{-1}L_{2}^{-}, \\
L_{2}^{-}\left( \partial ^{2}\right) ^{-1} &=&q^{-1}\left( \partial
^{2}\right) ^{-1}L_{2}^{-},  \nonumber \\
L_{2}^{-}\left( \partial ^{3}\right) ^{-1} &=&q\left( \partial ^{3}\right)
^{-1}L_{2}^{-}+q^{3}\partial ^{1}\left( \partial ^{3}\right) ^{-2}, 
\nonumber \\
L_{2}^{-}\left( \partial ^{4}\right) ^{-1} &=&q\left( \partial ^{4}\right)
^{-1}L_{2}^{-}-q^{3}\partial ^{2}\left( \partial ^{4}\right) ^{-2}, 
\nonumber \\[0.4cm]
K_{1}\left( \partial ^{1}\right) ^{-1} &=&q\left( \partial ^{1}\right)
^{-1}K_{1}, \\
K_{1}\left( \partial ^{2}\right) ^{-1} &=&q^{-1}\left( \partial ^{2}\right)
^{-1}K_{1},  \nonumber \\
K_{1}\left( \partial ^{3}\right) ^{-1} &=&q\left( \partial ^{3}\right)
^{-1}K_{1},  \nonumber \\
K_{1}\left( \partial ^{4}\right) ^{-1} &=&q^{-1}\left( \partial ^{4}\right)
^{-1}K_{1},  \nonumber \\[0.4cm]
K_{2}\left( \partial ^{1}\right) ^{-1} &=&q\left( \partial ^{1}\right)
^{-1}K_{2}, \\
K_{2}\left( \partial ^{2}\right) ^{-1} &=&q\left( \partial ^{2}\right)
^{-1}K_{2},  \nonumber \\
K_{2}\left( \partial ^{3}\right) ^{-1} &=&q^{-1}\left( \partial ^{3}\right)
^{-1}K_{2},  \nonumber \\
K_{2}\left( \partial ^{4}\right) ^{-1} &=&q^{-1}\left( \partial ^{4}\right)
^{-1}K_{2}.\nonumber
\end{eqnarray}
Applying the substitutions 
\begin{equation}
\partial ^{i}\rightarrow \hat{\partial}^{i},\quad (\partial
^{i})^{-1}\rightarrow (\hat{\partial}^{i})^{-1},\quad i=1,\ldots ,4,
\end{equation}
we get the corresponding relations for the second differential calculus
spanned by the conjugated derivatives.

With the representations in \cite{BW01} the same reasonings we have already
applied to the 3-dimensional Euclidean space lead immediately to the
expressions 
\begin{eqnarray}
(\hat{\partial}^{1})^{-1}\triangleright f &=&q\sum_{k=0}^{\infty }(-\lambda
)^{k}q^{-k(k+1)}\left[
x^{1}(D_{q^{-2}}^{4})^{-1}D_{q^{-2}}^{2}D_{q^{-2}}^{3}\right] ^{k} \\
&&\cdot\, (D_{q^{-2}}^{4})^{-1}f(q^{k+1}x^{2},q^{k+1}x^{3}),  \nonumber \\
(\hat{\partial}^{2})^{-1}\triangleright f &=&(D_{q^{-2}}^{3})^{-1}f(qx^{1}),
\nonumber \\
(\hat{\partial}^{3})^{-1}\triangleright f &=&(D_{q^{-2}}^{2})^{-1}f(qx^{1}),
\nonumber \\
(\hat{\partial}^{4})^{-1}\triangleright f &=&q^{-1}(D_{q^{-2}}^{1})^{-1}f. 
\nonumber
\end{eqnarray}
In complete analogy to the 3-dimensional case we have the transformation 
\begin{equation}
(\partial ^{i})^{-1}\,\tilde{\triangleright}\,
f\stackrel{{{\QATOP{i}{q}}{\QATOP{%
\rightarrow }{\rightarrow }}{\QATOP{i^{\prime }}{1/q}}}}{\longleftrightarrow 
}(\hat{\partial}^{i^{\prime }})^{-1}\triangleright f,\quad i^{\prime }=5-i,
\label{TrafLiReInt4dim}
\end{equation}
whose explicit form is given by the substitutions 
\begin{equation}
x^{\pm }\rightarrow x^{\mp },\quad q^{\pm 1}\rightarrow q^{\mp 1},\quad \hat{%
n}^{i}\rightarrow -\hat{n}^{i^{\prime }}
\end{equation}
\[
D_{q^{a}}^{i}\rightarrow D_{q^{-a}}^{i^{\prime }},\quad
(D_{q^{a}}^{i})^{-1}\rightarrow (D_{q^{-a}}^{i^{\prime }})^{-1}. 
\]
For the relationship between right and left integrals we now have 
\begin{eqnarray}
f\triangleleft (\partial ^{i})^{-1} &\stackrel{j\leftrightarrow j^{\prime }%
}{\longleftrightarrow }&-q^{4}(\hat{\partial}^{i^{\prime
}})^{-1}\triangleright f,  \label{TrafLiReCon4dim} \\
f\,\tilde{\triangleleft}\,(\hat{\partial}^{i})^{-1}
 &\stackrel{j\leftrightarrow
j^{\prime }}{\longleftrightarrow }&-q^{-4}(\partial ^{i^{\prime }})^{-1}%
\,\tilde{\triangleright}\,f,  \nonumber
\end{eqnarray}
which symbolizes that we have to apply the substitutions 
\begin{equation}
x^{j}\rightarrow x^{j^{\prime }},\quad D_{q^{a}}^{j}\rightarrow
D_{q^{a}}^{j^{\prime }},\quad (D_{q^{a}}^{j})^{-1}\rightarrow
(D_{q^{a}}^{j^{\prime }})^{-1},\quad \hat{n}^{j}\rightarrow \hat{n}%
^{j^{\prime }}.
\end{equation}

Next, we would like to present the formulae for integration by parts. For
left integrals belonging to the differential calculus with unhatted
derivatives we find 
\begin{eqnarray}
&&\left. (\partial ^{1})_{L}^{-1}(\partial _{L}^{1}f)\star g\right|
_{x^{4}=0}^{a} \\
&=&f\star g\left\| _{x^{4}=0}^{a}\right. -{}\left. (\partial
^{1})_{L}^{-1}(\Lambda ^{1/2}K_{1}^{1/2}K_{2}^{1/2}f)\star \partial
_{L}^{1}g\right| _{x^{4}=0}^{a},  \nonumber \\[0.16in]
&&\left. (\partial ^{2})_{L}^{-1}(\partial _{L}^{2}f)\star g\right|
_{x^{3}=0}^{a} \\
&=&f\star g\left\| _{x^{3}=0}^{a}\right. -\left. (\partial
^{2})_{L}^{-1}(\Lambda ^{1/2}K_{1}^{-1/2}K_{2}^{1/2}f)\star \partial
_{L}^{2}g\right| _{x^{3}=0}^{a}  \nonumber \\
&&-\,q\lambda \left. (\partial ^{2})_{L}^{-1}(\Lambda
^{1/2}K_{1}^{1/2}K_{2}^{1/2}L_{1}^{+}f)\star \partial _{L}^{1}g\right|
_{x^{3}=0}^{a},  \nonumber \\[0.16in]
&&\left. (\partial ^{3})_{L}^{-1}(\partial _{L}^{3}f)\star g\right|
_{x^{2}=0}^{a} \\
&=&f\star g\left\| _{x^{2}=0}^{a}\right. -\left. (\partial
^{3})_{L}^{-1}(\Lambda ^{1/2}K_{1}^{1/2}K_{2}^{-1/2}f)\star \partial
_{L}^{3}g\right| _{x^{3}=0}^{a}  \nonumber \\
&&-\,q\lambda \left. (\partial ^{3})_{L}^{-1}(\Lambda
^{1/2}K_{1}^{1/2}K_{2}^{1/2}L_{2}^{+}f)\star \partial _{L}^{1}g\right|
_{x^{2}=0}^{a},  \nonumber \\[0.16in]
&&\left. (\partial ^{4})_{L}^{-1}(\partial _{L}^{4}f)\star g\right|
_{x^{1}=a}^{b} \\
&=&f\star g\left\| _{x^{1}=0}^{a}\right. -\left. (\partial
^{4})_{L}^{-1}(\Lambda ^{1/2}K_{1}^{-1/2}K_{2}^{-1/2}f)\star \partial
_{L}^{4}g\right| _{x^{1}=a}^{b}  \nonumber \\
&&+\,q\lambda \left. (\partial ^{4})_{L}^{-1}(\Lambda
^{1/2}K_{1}^{1/2}K_{2}^{-1/2}L_{1}^{+}f)\star \partial _{L}^{3}g\right|
_{x^{1}=a}^{b}  \nonumber \\
&&+\,q\lambda \left. (\partial ^{4})_{L}^{-1}(\Lambda
^{1/2}K_{1}^{-1/2}K_{2}^{1/2}L_{2}^{+}f)\star \partial _{L}^{2}g\right|
_{x^{1}=a}^{b}  \nonumber \\
&&+\,q^{2}\lambda ^{2}\left. (\partial ^{4})_{L}^{-1}(\Lambda
^{1/2}K_{1}^{1/2}K_{2}^{1/2}L_{1}^{+}L_{2}^{+}f)\star \partial
_{L}^{1}g\right| _{x^{1}=a}^{b}.  \nonumber
\end{eqnarray}
For the second differential calculus with hatted derivatives one can
compute likewise 
\begin{eqnarray}
&&\left. (\hat{\partial}^{1})_{L}^{-1}(\hat{\partial}_{L}^{1}f)\star
g\right| _{x^{4}=0}^{a} \\
&=&f\star g\left\| _{x^{4}=0}^{a}\right. -{}\left. (\hat{\partial}%
^{1})_{L}^{-1}(\Lambda ^{-1/2}K_{1}^{-1/2}K_{2}^{-1/2}f)\star \hat{\partial}%
_{L}^{1}g\right| _{x^{4}=0}^{a}  \nonumber \\
&&+\,q^{-1}\lambda \left. (\hat{\partial}^{1})_{L}^{-1}(\Lambda
^{-1/2}K_{1}^{1/2}K_{2}^{-1/2}L_{1}^{-}f)\star \hat{\partial}%
_{L}^{2}g\right| _{x^{4}=0}^{a}  \nonumber \\
&&+\,q^{-1}\lambda \left. (\hat{\partial}^{1})_{L}^{-1}(\Lambda
^{-1/2}K_{1}^{-1/2}K_{2}^{1/2}L_{2}^{-}f)\star \hat{\partial}%
_{L}^{3}g\right| _{x^{4}=0}^{a}  \nonumber \\
&&+\,q^{-2}\lambda ^{2}\left. (\hat{\partial}^{1})_{L}^{-1}(\Lambda
^{-1/2}K_{1}^{1/2}K_{2}^{-1/2}L_{1}^{-}L_{2}^{-}f)\star \hat{\partial}%
_{L}^{4}g\right| _{x^{4}=0}^{a},  \nonumber \\[0.16in]
&&\left. (\hat{\partial}^{2})_{L}^{-1}(\hat{\partial}_{L}^{2}f)\star
g\right| _{x^{3}=a}^{b} \\
&=&f\star g\left\| _{x^{3}=0}^{a}\right. -\left. (\hat{\partial}%
^{2})_{L}^{-1}(\Lambda ^{-1/2}K_{1}^{1/2}K_{2}^{-1/2}f)\star \hat{\partial}%
_{L}^{2}g\right| _{x^{3}=0}^{a}  \nonumber \\
&&-\,q^{-1}\lambda \left. (\hat{\partial}^{2})_{L}^{-1}(\Lambda
^{-1/2}K_{1}^{1/2}K_{2}^{1/2}L_{2}^{-}f)\star \hat{\partial}_{L}^{4}g\right|
_{x^{3}=0}^{a},  \nonumber \\[0.16in]
&&\left. (\hat{\partial}^{3})_{L}^{-1}(\hat{\partial}_{L}^{3}f)\star g\right|
\\
&=&f\star g\left\| _{x^{2}=0}^{a}\right. -\left. (\hat{\partial}%
^{3})_{L}^{-1}(\Lambda ^{-1/2}K_{1}^{-1/2}K_{2}^{1/2}f)\star \hat{\partial}%
_{L}^{3}g\right| _{x^{2}=0}^{a}  \nonumber \\
&&-\,q^{-1}\lambda \left. (\hat{\partial}^{3})_{L}^{-1}(\Lambda
^{-1/2}K_{1}^{1/2}K_{2}^{1/2}L_{1}^{-}f)\star \hat{\partial}_{L}^{4}g\right|
_{x^{2}=0}^{a},  \nonumber \\[0.16in]
&&\left. (\hat{\partial}^{4})_{L}^{-1}(\hat{\partial}_{L}^{4}f)\star
g\right| _{x^{1}=a}^{b} \\
&=&f\star g\left\| _{x^{1}=0}^{a}\right. -\left. (\hat{\partial}%
^{4})_{L}^{-1}(\Lambda ^{-1/2}K_{1}^{1/2}K_{2}^{1/2}f)\star \hat{\partial}%
_{L}^{4}g\right| _{x^{1}=0}^{a}.  \nonumber
\end{eqnarray}
Now, we come to the corresponding formulae for right integrals which in the
case of the differential calculus with unhated derivatives read 
\begin{eqnarray}
&&\left. (\partial ^{1})_{R}^{-1}f\star \partial _{R}^{1}g\right|
_{x^{4}=0}^{a} \\
&=&f\star g\left\| _{x^{4}=0}^{a}\right. -\left. (\partial
^{1})_{R}^{-1}(\partial _{R}^{1}f)\star (\Lambda
^{-1/2}K_{1}^{-1/2}K_{2}^{-1/2}g)\right| _{x^{4}=0}^{a},  \nonumber \\%
[0.16in]
&&\left. (\partial ^{2})_{R}^{-1}f\star \partial _{R}^{2}g\right|
_{x^{3}=0}^{a} \\
&=&f\star g\left\| _{x^{3}=0}^{a}\right. -\left. (\partial
^{2})_{R}^{-1}(\partial _{R}^{2}f)\star (\Lambda
^{-1/2}K_{1}^{1/2}K_{2}^{-1/2}g)\right| _{x^{3}=0}^{a}  \nonumber \\
&&+\,\lambda \left. (\partial ^{2})_{R}^{-1}(\partial _{R}^{1}f)\star
(\Lambda ^{-1/2}K_{1}^{1/2}K_{2}^{-1/2}L_{1}^{+}g)\right| _{x^{3}=0}^{a}, 
\nonumber \\[0.16in]
&&\left. (\partial ^{3})_{R}^{-1}f\star \partial _{R}^{3}g\right|
_{x^{2}=0}^{a} \\
&=&f\star g\left\| _{x^{2}=0}^{a}\right. -\left. (\partial
^{3})_{R}^{-1}(\partial _{R}^{3}f)\star (\Lambda
^{-1/2}K_{1}^{-1/2}K_{2}^{1/2}g)\right| _{x^{2}=0}^{a}  \nonumber \\
&&+\,\lambda \left. (\partial ^{3})_{R}^{-1}(\partial _{R}^{1}f)\star
(\Lambda ^{-1/2}K_{1}^{-1/2}K_{2}^{1/2}L_{2}^{+}g)\right| _{x^{2}=0}^{a}, 
\nonumber \\[0.16in]
&&\left. (\partial ^{4})_{R}^{-1}f\star \partial _{R}^{4}g\right|
_{x^{1}=0}^{a} \\
&=&f\star g\left\| _{x^{1}=0}^{a}\right. -\left. (\partial
^{4})_{R}^{-1}(\partial _{R}^{4}f)\star (\Lambda
^{-1/2}K_{1}^{1/2}K_{2}^{1/2}g)\right| _{x^{1}=0}^{a}  \nonumber \\
&&-\,\lambda \left. (\partial ^{4})_{R}^{-1}(\partial _{R}^{3}f)\star
(\Lambda ^{-1/2}K_{1}^{1/2}K_{2}^{1/2}L_{1}^{+}g)\right| _{x^{1}=0}^{a} 
\nonumber \\
&&-\,\lambda \left. (\partial ^{4})_{R}^{-1}(\partial _{R}^{2}f)\star
(\Lambda ^{-1/2}K_{1}^{1/2}K_{2}^{1/2}L_{2}^{+}g)\right| _{x^{1}=0}^{a} 
\nonumber \\
&&+\,\lambda ^{2}\left. (\partial ^{4})_{R}^{-1}(\partial _{R}^{1}f)\star
(\Lambda ^{-1/2}K_{1}^{1/2}K_{2}^{1/2}L_{1}^{+}L_{2}^{+}g)\right|
_{x^{1}=0}^{a}.  \nonumber
\end{eqnarray}
In the same manner we get for the second differential calculus with
unhated derivatives 
\begin{eqnarray}
&&\left. (\hat{\partial}^{1})_{R}^{-1}f\star \hat{\partial}_{R}^{1}g\right|
_{x^{4}=0}^{a} \\
&=&f\star g\left\| _{x^{4}=0}^{a}\right. -\left. (\hat{\partial}%
^{1})_{R}^{-1}(\hat{\partial}_{R}^{1}f)\star (\Lambda
^{1/2}K_{1}^{1/2}K_{2}^{1/2}g)\right| _{x^{4}=a}^{b}  \nonumber \\
&&-\,\lambda \left. (\hat{\partial}^{1})_{R}^{-1}(\hat{\partial}%
_{R}^{2}f)\star (\Lambda ^{1/2}K_{1}^{1/2}K_{2}^{1/2}L_{1}^{-}g)\right|
_{x^{4}=a}^{b}  \nonumber \\
&&-\,\lambda \left. (\hat{\partial}^{1})_{R}^{-1}(\hat{\partial}%
_{R}^{3}f)\star (\Lambda ^{1/2}K_{1}^{1/2}K_{2}^{1/2}L_{2}^{-}g)\right|
_{x^{4}=a}^{b}  \nonumber \\
&&+\,\lambda ^{2}\left. (\hat{\partial}^{1})_{R}^{-1}(\hat{\partial}%
_{R}^{4}f)\star (\Lambda
^{1/2}K_{1}^{1/2}K_{2}^{1/2}L_{1}^{-}L_{2}^{-}g)\right| _{x^{4}=a}^{b}, 
\nonumber \\[0.16in]
&&\left. (\hat{\partial}^{2})_{R}^{-1}f\star \hat{\partial}_{R}^{2}g\right|
_{x^{3}=0}^{a} \\
&=&f\star g\left\| _{x^{3}=0}^{a}\right. -\left. (\hat{\partial}%
^{2})_{R}^{-1}(\hat{\partial}_{R}^{2}f)\star (\Lambda
^{1/2}K_{1}^{-1/2}K_{2}^{1/2}g)\right| _{x^{3}=0}^{a}  \nonumber \\
&&+\,\lambda \left. (\hat{\partial}^{2})_{R}^{-1}(\hat{\partial}%
_{R}^{4}f)\star (\Lambda ^{1/2}K_{1}^{-1/2}K_{2}^{1/2}L_{2}^{-}g)\right|
_{x^{3}=0}^{a},  \nonumber \\[0.16in]
&&\left. (\hat{\partial}^{3})_{R}^{-1}f\star \hat{\partial}_{R}^{3}g\right|
_{x^{2}=0}^{a} \\
&=&f\star g\left\| _{x^{2}=0}^{a}\right. -\left. (\hat{\partial}%
^{3})_{R}^{-1}(\hat{\partial}_{R}^{3}f)\star (\Lambda
^{1/2}K_{1}^{-1/2}K_{2}^{-1/2}g)\right| _{x^{2}=0}^{a}  \nonumber \\
&&+\,\lambda \left. (\hat{\partial}^{3})_{R}^{-1}(\hat{\partial}%
_{R}^{4}f)\star (\Lambda ^{1/2}K_{1}^{1/2}K_{2}^{-1/2}L_{1}^{-}g)\right|
_{x^{2}=0}^{a},  \nonumber \\[0.16in]
&&\left. (\hat{\partial}^{4})_{R}^{-1}f\star \hat{\partial}_{R}^{4}g\right|
_{x^{1}=a}^{b} \\
&=&f\star g\left\| _{x^{1}=0}^{a}\right. -\left. (\hat{\partial}%
^{4})_{R}^{-1}(\hat{\partial}_{R}^{4}f)\star (\Lambda
^{1/2}K_{1}^{-1/2}K_{2}^{-1/2}g)\right| _{x^{1}=0}^{a}.  \nonumber
\end{eqnarray}

Finally, we would like to present the relations concerning our method for
integration over the whole space. If surface terms are neglected, we can
again state the identities 
\begin{eqnarray}
(\partial ^{1})^{-1}(\partial ^{2})^{-1}(\partial ^{3})^{-1}(\partial
^{4})^{-1} &=&(\partial ^{1})^{-1}(\partial ^{3})^{-1}(\partial
^{2})^{-1}(\partial ^{4})^{-1}=  \label{volume4} \\
q(\partial ^{1})^{-1}(\partial ^{2})^{-1}(\partial ^{4})^{-1}(\partial
^{3})^{-1} &=&q(\partial ^{1})^{-1}(\partial ^{3})^{-1}(\partial
^{4})^{-1}(\partial ^{2})^{-1}=  \nonumber \\
q(\partial ^{2})^{-1}(\partial ^{1})^{-1}(\partial ^{3})^{-1}(\partial
^{4})^{-1} &=&q(\partial ^{3})^{-1}(\partial ^{1})^{-1}(\partial
^{2})^{-1}(\partial ^{4})^{-1}=  \nonumber \\
q^{3}(\partial ^{2})^{-1}(\partial ^{4})^{-1}(\partial ^{3})^{-1}(\partial
^{1})^{-1} &=&q^{3}(\partial ^{3})^{-1}(\partial ^{4})^{-1}(\partial
^{2})^{-1}(\partial ^{1})^{-1}=  \nonumber \\
q^{3}(\partial ^{4})^{-1}(\partial ^{2})^{-1}(\partial ^{1})^{-1}(\partial
^{3})^{-1} &=&q^{3}(\partial ^{4})^{-1}(\partial ^{3})^{-1}(\partial
^{1})^{-1}(\partial ^{2})^{-1}=  \nonumber \\
q^{4}(\partial ^{4})^{-1}(\partial ^{2})^{-1}(\partial ^{3})^{-1}(\partial
^{1})^{-1} &=&q^{4}(\partial ^{4})^{-1}(\partial ^{3})^{-1}(\partial
^{2})^{-1}(\partial ^{1})^{-1}=  \nonumber \\
&=&q^{2}\;\times \text{remaining combinations.}  \nonumber
\end{eqnarray}
Let us note that the quantities $(\hat{\partial}^{i})^{-1}$ $,$ $i=1,\ldots
,4,$ obey the same relations. Since the results of the various volume
integrals in (\ref{volume4}) differ by normalisation factors only, we can
restrict attention to one of the above expressions. Thus, for performing the
integration over the whole space it is sufficient to consider the following
formula: 
\begin{eqnarray}
&&(\hat{\partial}^{1})^{-1}(\hat{\partial}^{2})^{-1}(\hat{\partial}%
^{3})^{-1}(\hat{\partial}^{4})^{-1}\triangleright f \\
&=&q^{4}(D_{q^{-2}}^{4})^{-1}(D_{q^{-2}}^{3})^{-1}(D_{q^{-2}}^{2})^{-1}(D_{q^{-2}}^{1})^{-1}f(q^{2}x^{1},qx^{2},qx^{3}).
\nonumber
\end{eqnarray}
It is quite clear that the transformations (\ref{TrafLiReInt4dim}) and (\ref
{TrafLiReCon4dim}) carry over into our formulae for volume integrals. Hence, 
\begin{eqnarray}
&&(\hat{\partial}^{1})^{-1}(\hat{\partial}^{2})^{-1}(\hat{\partial}%
^{3})^{-1}(\hat{\partial}^{4})^{-1}\triangleright f \\
&\stackrel{{{\QATOP{i}{q}}{\QATOP{\rightarrow }{\rightarrow }}{\QATOP{%
i^{\prime }}{1/q}}}}{\longleftrightarrow } &(\partial ^{4})^{-1}(\partial
^{3})^{-1}(\partial ^{2})^{-1}(\partial ^{1})^{-1}\,\tilde{\triangleright}\,f 
\nonumber
\end{eqnarray}
and 
\begin{eqnarray}
&&f\,\tilde{\triangleleft}\,
(\bar{\partial}^{1})^{-1}(\bar{\partial}^{2})^{-1}(%
\bar{\partial}^{3})^{-1}(\bar{\partial}^{4})^{-1} \\
&\stackrel{j\leftrightarrow j^{\prime }}{\longleftrightarrow } &(\partial
^{4})^{-1}(\partial ^{3})^{-1}(\partial ^{2})^{-1}(\partial ^{1})^{-1}
\,\tilde{%
\triangleright}\,f,  \nonumber \\[0.16in]
&&f\triangleleft (\partial ^{4})^{-1}(\partial ^{3})^{-1}(\partial
^{2})^{-1}(\partial ^{1})^{-1} \\
&\stackrel{j\leftrightarrow j^{\prime }}{\longleftrightarrow } &(\bar{%
\partial}^{1})^{-1}(\bar{\partial}^{2})^{-1}(\bar{\partial}^{3})^{-1}(\bar{%
\partial}^{4})^{-1}\triangleright f,  \nonumber
\end{eqnarray}
where $(\partial ^{i})^{-1}=-q^{4}(\hat{\partial}^{i})^{-1}.$ With the same
reasonings applied to q-deformed Euclidean space in three dimensions we can
immediately verify rotation and translation invariance of our 
4-dimensional volume integrals. Using the notation 
\begin{eqnarray}
\int_{L}d_{q}V{}f &\equiv &(\partial ^{4})^{-1}(\partial ^{3})^{-1}(\partial
^{2})^{-1}(\partial ^{1})^{-1}\,\tilde{\triangleright}\,f, \\
\int_{R}d_{q}V{}f &\equiv &f\triangleleft (\partial ^{4})^{-1}(\partial
^{3})^{-1}(\partial ^{2})^{-1}(\partial ^{1})^{-1},  \nonumber \\
\int_{L}d_{q}\overline{V}{}f &\equiv &(\bar{\partial}^{1})^{-1}(\bar{\partial%
}^{2})^{-1}(\bar{\partial}^{3})^{-1}(\bar{\partial}^{4})^{-1}\triangleright
f,  \nonumber \\
\int_{R}d_{q}\overline{V}{}f &\equiv &f\,\tilde{\triangleleft}\,
(\bar{\partial}%
^{1})^{-1}(\bar{\partial}^{2})^{-1}(\bar{\partial}^{3})^{-1}(\bar{\partial}%
^{4})^{-1},  \nonumber
\end{eqnarray}
we can explicitely write 
\begin{eqnarray}
\partial ^{i}\,\tilde{\triangleright} \int_{L/R}d_{q}V{}f
&=&\int_{L/R}d_{q}V{}\partial ^{i}\,\tilde{\triangleright}\,
 f=\varepsilon (\partial
^{i})\int_{L/R}d_{q}V{}f=0, \\
L_{j}^{\pm }\triangleright \int_{L/R}d_{q}V{}f
&=&\int_{L/R}d_{q}V{}L_{j}^{\pm }\triangleright f=\varepsilon (L_{j}^{\pm
})\int_{L/R}d_{q}V{}f=0,  \nonumber \\
K_{j}\triangleright \int_{L/R}d_{q}V{}f
&=&\int_{L/R}d_{q}V{}K_{j}\triangleright f=\varepsilon
(K_{j})\int_{L/R}d_{q}V{}f=\int_{L/R}d_{q}V{}f,  \nonumber \\[0.16in]
\bar{\partial}^{i}\triangleright \int_{L/R}d_{q}\overline{V}{}f
&=&\int_{L/R}d_{q}\overline{V}{}\bar{\partial}^{i}\triangleright
f=\varepsilon (\bar{\partial}^{i})\int_{L/R}d_{q}\overline{V}{}f=0, \\
L_{j}^{\pm }\triangleright \int_{L/R}d_{q}\overline{V}{}f &=&\int_{L/R}d_{q}%
\overline{V}{}L_{j}^{\pm }\triangleright f=\varepsilon (L_{j}^{\pm
})\int_{L/R}d_{q}\overline{V}{}f=0,  \nonumber \\
K_{j}\triangleright \int_{L/R}d_{q}\overline{V}{}f &=&\int_{L/R}d_{q}%
\overline{V}{}K_{j}\triangleright f=\varepsilon (K_{j})\int_{L/R}d_{q}%
\overline{V}{}f=\int_{L/R}d_{q}\overline{V}{}f  \nonumber
\end{eqnarray}
and 
\begin{eqnarray}
\left( \int_{L/R}d_{q}V{}f\right) \,\tilde{\triangleleft}\, \partial ^{i}
&=&\int_{L/R}d_{q}V{}f\,\tilde{\triangleleft}\, \partial ^{i}=\varepsilon (\partial
^{i})\int_{L/R}d_{q}V{}f=0, \\
\left( \int_{L/R}d_{q}V{}f\right) \triangleleft L_{j}^{\pm }
&=&\int_{L/R}d_{q}V{}f\triangleleft L_{j}^{\pm }=\varepsilon (L_{j}^{\pm
})\int_{L/R}d_{q}V{}f=0,  \nonumber \\
\left( \int_{L/R}d_{q}V{}f\right) \triangleleft K_{j}
&=&\int_{L/R}d_{q}V{}f\triangleleft K_{j}=\varepsilon
(K_{j})\int_{L/R}d_{q}V{}f=\int_{L/R}d_{q}V{}f,  \nonumber \\[0.16in]
\left( \int_{L/R}d_{q}\overline{V}{}f\right) \triangleleft \bar{\partial}%
^{i} &=&\int_{L/R}d_{q}\overline{V}{}f\triangleleft \bar{\partial}%
^{i}=\varepsilon (\bar{\partial}^{i})\int_{L/R}d_{q}V{}f=0, \\
\left( \int_{L/R}d_{q}\overline{V}{}f\right) \triangleleft L_{j}^{\pm }
&=&\int_{L/R}d_{q}\overline{V}{}f\triangleleft L_{j}^{\pm }=\varepsilon
(L_{j}^{\pm })\int_{L/R}d_{q}\overline{V}{}f=0,  \nonumber \\
\left( \int_{L/R}d_{q}\overline{V}{}f\right) \triangleleft K_{j}
&=&\int_{L/R}d_{q}\overline{V}{}f\triangleleft K_{j}=\varepsilon
(K_{j})\int_{L/R}d_{q}\overline{V}{}f=\int_{L/R}d_{q}\overline{V}{}f. 
\nonumber
\end{eqnarray}

\section{q-Deformed Minkowski-space\label{KapMin}}

In principle all considerations of the previous two sections pertain equally
to q-deformed Minkowski space \cite{LWW97} \cite{CSSW90}, \cite{SWZ91}, \cite
{Maj91}\footnote{%
For a different version of q-deformed Minkowski space see also \cite{Dob94}.},
apart from the fact that the results now entail a more involved structure. The
partial derivatives of q-deformed Minkowski space satisfy the relations 
\begin{eqnarray}
\partial ^{\mu }\partial ^{0} &=&\partial ^{0}\partial ^{\mu },\quad \mu
=\{0,+,-,3/0\},  \label{Derivative} \\
\partial ^{-}\partial ^{3/0} &=&q^{2}\partial ^{3/0}\partial ^{-},  \nonumber
\\
\partial ^{+}\partial ^{3/0} &=&q^{-2}\partial ^{3/0}\partial ^{+}, 
\nonumber \\
\partial ^{-}\partial ^{+}-\partial ^{+}\partial ^{-} &=&\lambda (\partial
^{3/0}\partial ^{3/0}+\partial ^{0}\partial ^{3/0}),  \nonumber
\end{eqnarray}
with $\lambda =q-q^{-1}$ and $q>1.$ As usual, we introduce inverse elements $%
(\partial ^{\mu })^{-1}$, $\mu =\pm ,0,3/0,$ by 
\begin{equation}
\left( \partial ^{\mu }\right) ^{-1}\partial ^{\mu }=\partial ^{\mu }\left(
\partial ^{\mu }\right) ^{-1}=1,\quad \mu =0,+,-,3/0.
\end{equation}
From these requirements we get the relations 
\begin{eqnarray}
\partial ^{0}(\partial ^{\mu })^{-1} &=&(\partial ^{\mu })^{-1}\partial
^{0},\quad \mu =0,+,-,3/0, \\
(\partial ^{3/0})^{-1}\partial ^{\pm } &=&q^{\mp 2}\partial ^{\pm }(\partial
^{3/0})^{-1},  \nonumber \\
(\partial ^{\pm })^{-1}\partial ^{3/0} &=&q^{\pm 2}\partial ^{3/0}(\partial
^{\pm })^{-1},  \nonumber \\
\partial ^{-}(\partial ^{+})^{-1} &=&(\partial ^{+})^{-1}\partial
^{-}-q^{-2}\lambda (\partial ^{+})^{-2}(q^{-2}\partial ^{3/0}+\partial
^{0})\partial ^{3/0},  \nonumber \\
(\partial ^{-})^{-1}\partial ^{+} &=&\partial ^{+}(\partial
^{-})^{-1}-q^{-2}\lambda \partial ^{3/0}(q^{-2}\partial ^{3/0}+\partial
^{0})(\partial ^{-})^{-2}.  \nonumber
\end{eqnarray}
Furthermore, we have 
\begin{eqnarray}
(\partial ^{0})^{-1}(\partial ^{\mu })^{-1} &=&(\partial ^{\mu
})^{-1}(\partial ^{0})^{-1},\quad \mu =0,+,-,3/0, \\
(\partial ^{3/0})^{-1}(\partial ^{\pm })^{-1} &=&q^{\pm 2}(\partial
^{+})^{-1}(\partial ^{3/0})^{-1},  \nonumber \\
(\partial ^{-})^{-1}(\partial ^{+})^{-1} &=&q^{2}\sum_{i=0}^{\infty
}(\lambda \lambda _{+}^{-1})^{i}\left[ \left[ i\right] \right]
_{q^{2}}!\left( \QATOPD {-1}{i}_{q^{-2}}\right) ^{2}  \nonumber \\
&&\cdot \sum_{j+k=i}(-q^{6})^{k}q^{i\left( i+2k\right) }\QATOPD {i}{k}%
_{q^{2}}\sum_{p=0}^{k}(q^{4j}\lambda _{+})^{p}  \nonumber \\
&&\cdot\, {}(\partial ^{+})^{p-\left( i+1\right) }(\partial
^{3/0})^{2j}(S_{q})_{k,p}(\partial ^{0},\partial ^{3/0})(\partial
^{-})^{p-\left( i+1\right) },  \nonumber
\end{eqnarray}
where $S_{k,p}$ is a polynomial of degree $2(k-p)$ with its explicit form
presented in appendix \ref{AppA}.

Next, we come to the commutation relations involving the Lorentz generators 
\cite{OSWZ92}, \cite{SWZ91}, \cite{RW99}. Explicitely, they are given by 
\begin{eqnarray}
T^{+}(\partial ^{0})^{-1} &=&(\partial ^{0})^{-1}T^{+}, \\
T^{+}(\partial ^{3/0})^{-1} &=&(\partial ^{3/0})^{-1}T^{+}-q^{1/2}\lambda
_{+}^{1/2}(\partial ^{3/0})^{-2}\partial ^{+},  \nonumber \\
T^{+}(\partial ^{+})^{-1} &=&q^{2}(\partial ^{+})^{-1}T^{+},  \nonumber \\
T^{+}(\partial ^{-})^{-1} &=&q^{-2}(\partial ^{-})^{-1}T^{+}-q^{-1/2}\lambda
_{+}^{1/2}(\partial ^{-})^{-2}(\partial ^{3/0}+q^{-2}\partial ^{0}), 
\nonumber \\[0.16in]
T^{-}(\partial ^{0})^{-1} &=&(\partial ^{0})^{-1}T^{-} \\
T^{-}(\partial ^{3/0})^{-1} &=&(\partial ^{3/0})^{-1}T^{-}-q^{-1/2}\lambda
_{+}^{1/2}(\partial ^{3/0})^{-2}\partial ^{-},  \nonumber \\
T^{-}(\partial ^{-})^{-1} &=&q^{-2}(\partial ^{-})^{-1}T^{-},  \nonumber \\
T^{-}(\partial ^{+})^{-1} &=&q^{2}(\partial ^{+})^{-1}T^{-}-q^{1/2}\lambda
_{+}^{1/2}(\partial ^{+})^{-2}(\partial ^{3/0}+q^{2}\partial ^{0}), 
\nonumber \\[0.16in]
\tau ^{3}(\partial ^{0})^{-1} &=&(\partial ^{0})^{-1}\tau ^{3}, \\
\tau ^{3}(\partial ^{3/0})^{-1} &=&(\partial ^{3/0})^{-1}\tau ^{3}, 
\nonumber \\
\tau ^{3}(\partial ^{+})^{-1} &=&q^{4}(\partial ^{+})^{-1}\tau ^{3}, 
\nonumber \\[0.16in]
\tau ^{3}(\partial ^{-})^{-1} &=&q^{-4}(\partial ^{-})^{-1}\tau ^{3}, 
\nonumber \\
T^{2}(\partial ^{3/0})^{-1} &=&q(\partial ^{3/0})^{-1}T^{2}, \\
T^{2}(\partial ^{+})^{-1} &=&q^{-1}(\partial ^{+})^{-1}T^{2},  \nonumber \\
T^{2}(\partial ^{-})^{-1} &=&q(\partial ^{-})^{-1}T^{2}-q^{-3/2}\lambda
_{+}^{-1/2}\partial ^{3/0}(\partial ^{-})^{-2}\tau ^{1},  \nonumber \\
T^{2}(\partial ^{3})^{-1} &=&(T^{2}\triangleright \left( \partial
^{3}\right) ^{-1})\tau ^{1}+(\sigma ^{2}\triangleright \left( \partial
^{3}\right) ^{-1})T^{2},  \nonumber \\[0.16in]
S^{1}(\partial ^{3/0})^{-1} &=&q^{-1}(\partial ^{3/0})^{-1}S^{1}, \\
S^{1}(\partial ^{-})^{-1} &=&q^{-1}(\partial ^{-})^{-1}S^{1},  \nonumber \\
S^{1}\left( \partial ^{+}\right) ^{-1} &=&q(\partial
^{+})^{-1}S^{1}+q^{-1/2}\lambda _{+}^{-1/2}(\partial ^{+})^{-2}\partial
^{3/0}\sigma ^{2},  \nonumber \\
S^{1}(\partial ^{3})^{-1} &=&(S^{1}\triangleright \left( \partial
^{3}\right) ^{-1})\sigma ^{2}+(\tau ^{1}\triangleright \left( \partial
^{3}\right) ^{-1})S^{1},  \nonumber \\[0.16in]
\tau ^{1}(\partial ^{3/0})^{-1} &=&q^{-1}(\partial ^{3/0})^{-1}\tau ^{1}, \\
\tau ^{1}(\partial ^{-})^{-1} &=&q(\partial ^{-})^{-1}\tau ^{1},  \nonumber
\\
\tau ^{1}(\partial ^{+})^{-1} &=&q^{-1}(\partial ^{+})^{-1}\tau
^{1}+q^{-1/2}\lambda _{+}^{-1/2}\lambda ^{2}(\partial ^{+})^{-2}\partial
^{3/0}T^{2},  \nonumber \\
\tau ^{1}(\partial ^{3})^{-1} &=&(\tau ^{1}\triangleright (\partial
^{3})^{-1})\tau ^{1}+\lambda ^{2}(S^{1}\triangleright (\partial
^{3})^{-1})T^{2},  \nonumber \\[0.16in]
\sigma ^{2}(\partial ^{3/0})^{-1} &=&q(\partial ^{3/0})^{-1}\sigma ^{2}, \\
\sigma ^{2}(\partial ^{+})^{-1} &=&q(\partial ^{+})^{-1}\sigma ^{2}, 
\nonumber \\
\sigma ^{2}(\partial ^{-})^{-1} &=&q^{-1}(\partial ^{-})^{-1}\sigma
^{2}-q^{1/2}\lambda _{+}^{-1/2}\lambda ^{2}(\partial ^{-})^{-2}\partial
^{3/0}S^{1},  \nonumber \\
\sigma ^{2}(\partial ^{3})^{-1} &=&(\sigma ^{2}\triangleright (\partial
^{3})^{-1})\sigma ^{2}+\lambda ^{2}(T^{2}\triangleright (\partial
^{3})^{-1})S^{1},  \nonumber
\end{eqnarray}
where we have to insert the actions 
\begin{eqnarray}
T^{2}\triangleright (\partial ^{3})^{-1} &=&q^{-3/2}\sum_{k=0}^{\infty
}(q^{2}\alpha _{0})^{k}\left( K_{-1}\right) _{1,q^{2}}^{\left( k,k+1\right) }
\\
&&\cdot \sum_{0\leq i+j\leq k}\lambda _{+}^{j-1/2}\binom{k}{i}(\partial
^{+})^{j+1}(a_{q}(q^{2j}\partial ^{3/0}))^{i}  \nonumber \\
&&\cdot\, {}(S_q)_{k-i,j}(\partial ^{0},\partial ^{3/0})(\partial
^{0}+2q^{2j+1}\lambda _{+}^{-1}\partial ^{3/0})^{-2\left( k+1\right)
}(\partial ^{-})^{j},  \nonumber \\[0.1in]
S^{1}\triangleright (\partial ^{3})^{-1} &=&-q^{-3/2}\sum_{k=0}^{\infty
}(q^{-2}\alpha _{0})^{k}\left( K_{-1}\right) _{1,q^{-2}}^{\left(
k,k+1\right) } \\
&&\cdot \sum_{0\leq i+j\leq k}\lambda _{+}^{j-1/2}\binom{k}{i}(\partial
^{+})^{j}(a_{q}(q^{2j}\partial ^{3/0}))^{i}  \nonumber \\
&&\cdot\, {}(S_q)_{k-i,j}(\partial ^{0},\partial ^{3/0})(\partial
^{0}+2q^{2j+1}\lambda _{+}^{-1}\partial ^{3/0})^{-2\left( k+1\right)
}(\partial ^{-})^{j+1},  \nonumber \\[0.1in]
\tau ^{1}\triangleright (\partial ^{3})^{-1} &=&q\sum_{k=0}^{\infty
}(q^{2}\alpha _{0})^{k}\left( K_{-1}\right) _{1,q^{2}}^{\left( k,k\right) }
\\
&&\cdot \sum_{0\leq i+j\leq k}\lambda _{+}^{j}\binom{k}{i}(\partial
^{+})^{j}(a_{q}(q^{2j}\partial ^{3/0}))^{i}  \nonumber \\
&&\cdot \,{}(S_q)_{k-i,j}(\partial ^{0},\partial ^{3/0})(\partial
^{0}+2q^{2j+1}\lambda _{+}^{-1}\partial ^{3/0})^{-2k-1}(\partial ^{-})^{j}, 
\nonumber \\[0.1in]
\sigma ^{2}\triangleright (\partial ^{3})^{-1} &=&q^{-1}\sum_{k=0}^{\infty
}(q^{-2}\alpha _{0})^{k}\left( K_{-1}\right) _{1,q^{-2}}^{\left( k,k\right) }
\label{Wirkungen} \\
&&\cdot \sum_{0\leq i+j\leq k}\lambda _{+}^{j}\binom{k}{i}(\partial
^{+})^{j}(a_{q^{-1}}(q^{2j}\partial ^{3/0}))^{i}  \nonumber \\
&&\cdot\, {}(S_q)_{k-i,j}(\partial ^{0},\partial ^{3/0})(\partial
^{0}+2q^{2j-1}\lambda _{+}^{-1}\partial ^{3/0})^{-2k-1}(\partial ^{-})^{j} 
\nonumber
\end{eqnarray}
with 
\begin{equation}
\alpha _{0}=-\frac{\lambda ^{2}}{\lambda _{+}^{2}}.
\end{equation}
Note that these expressions have been formulated by using the abbreviations
and conventions listed in appendix \ref{AppA}. Applying the substitutions 
\begin{equation}
\partial ^{\mu }\rightarrow \hat{\partial}^{\mu },\qquad (\partial ^{\mu
})^{-1}\rightarrow (\hat{\partial}^{\mu })^{-1},\quad \mu =\pm ,0,3/0,
\end{equation}
to the formulae (\ref{Derivative})-(\ref{Wirkungen}) yields the
corresponding identities for the differential calculus of the hatted
derivatives.

As in the Euclidean case the representations of the partial derivatives of
q-deformed Minkowski space split into two parts \cite{BW01}. Hence 
\begin{equation}
\hat{\partial}^{\mu }\triangleright F=\left( (\hat{\partial}_{(i=0)}^{\mu
})+(\hat{\partial}_{\left( i>0\right) }^{\mu })\right) F,\quad \mu =\pm
,0,3/0.
\end{equation}
Now, we can follow the same lines as in the previous sections. Thus,
solutions to the difference equations 
\begin{equation}
\hat{\partial}^{\mu }\triangleright F=f,\quad \mu =\pm ,0,3/0
\end{equation}
are given by 
\begin{eqnarray}
F &=&(\hat{\partial}^{\mu })^{-1}\triangleright f  \label{IntegralM} \\
&=&\sum_{k=0}^{\infty }(-1)^{k}\left[ (\hat{\partial}_{(i=0)}^{\mu })^{-1}(%
\hat{\partial}_{(i>0)}^{\mu })\right] ^{k}(\hat{\partial}_{(i=0)}^{\mu
})^{-1}f,\quad \mu =\pm ,0,3/0.  \nonumber
\end{eqnarray}
The operators we have to insert in (\ref{IntegralM}) for evaluating can in
explicit terms be written as 
\begin{eqnarray}
(\hat{\partial}_{\left( i=0\right)
}^{3/0})_{L}^{-1}f(x^{+},x^{3/0},x^{3},x^{-})
&=&(D_{q_{-}q^{2}}^{3})^{-1}f(q^{-2}x^{+}),  \label{DarDerMinAnf} \\
(\hat{\partial}_{\left( i=0\right)
}^{-})_{L}^{-1}f(x^{+},x^{3/0},x^{3},x^{-}) &=&-q(D_{q^{2}}^{+})^{-1}f, 
\nonumber \\
(\hat{\partial}_{\left( i=0\right)
}^{+})_{L}^{-1}f(x^{+},x^{3/0},x^{3},x^{-})
&=&-q^{-1}(D_{q^{2}}^{-})^{-1}f(q^{-2}q_{+}x^{3}),  \nonumber \\
(\hat{\partial}_{\left( i=0\right)
}^{0})_{L}^{-1}f(x^{+},x^{3/0},x^{3},x^{-})
&=&(D_{q^{2}}^{3/0})^{-1}f(q_{-}x^{3})  \nonumber
\end{eqnarray}
and 
\begin{eqnarray}
&&(\hat{\partial}_{\left( i>0\right) }^{3/0})_{L}f(x^{+},x^{3/0},x^{3},x^{-})
\\
&=&\sum_{l=1}^{\infty }\alpha _{+}^{l}\sum_{0\leq i+j\leq l}\left(
M^{-}\right) _{i,j}^{l}(\underline{x})(T^{3})_{j}^{i}f,  \nonumber \\[0.16in]
&&(\hat{\partial}_{\left( i>0\right) }^{-})_{L}f(x^{+},x^{3/0},x^{3},x^{-})
\\
&=&\frac{\lambda }{\lambda _{+}}\sum_{l=0}^{\infty }\alpha
_{+}^{l}\sum_{0\leq i+j\leq l}\Big\{ (M^{+})_{i,j}^{k}(\underline{x}%
)(T_{1}^{-})_{j}^{i}f+q^{-1}(M^{-})_{i,j}^{k}(\underline{x}%
)(T_{2}^{-})_{j}^{i}f\Big\} ,  \nonumber \\[0.16in]
&&(\hat{\partial}_{\left( i>0\right) }^{+})_{L}f(x^{+},x^{3},x^{3/0},x^{-})
\label{DarDerMinEnd} \\
&=&-q\lambda x^{+}(D_{q^{2}}^{3/0}D_{q^{2}}^{3}f)^{-1}f(q_{+}x^{3}) 
\nonumber \\
&&-{\,}q\sum_{k=1}^{\infty }\alpha _{+}^{k}\sum_{0\leq i+j\leq k}\Big\{
(M^{-})_{i,j}^{k}(\underline{x})(T_{1}^{+})_{j}^{i}f+\lambda
(M^{+})_{i,j}^{k}(\underline{x})(T_{2}^{+})_{j}^{i}f \Big\}  \nonumber \\
&&-{\,}q\frac{\lambda }{\lambda _{+}}\sum_{0\leq k+l<\infty }\alpha
_{+}^{k+l}\sum_{i=0}^{k}\sum_{j=0}^{l}\sum_{0\leq u\leq
i+j}(M^{+-})_{i,j,u}^{k,l}(\underline{x})(T_{3}^{+})_{u}^{k,l}f  \nonumber \\
&&-{\,}\frac{\lambda }{\lambda _{+}}\sum_{0\leq k+l<\infty }\alpha
_{+}^{k+l+1}\sum_{i=0}^{k}\sum_{j=0}^{l+1}\sum_{0\leq u\leq
i+j}(M^{+-})_{i,j,u}^{k,l+1}(\underline{x})(T_{4}^{+})_{u}^{k,l}f,  \nonumber
\\ [0.16in]
&&(\hat{\partial}_{\left( i>0\right) }^{0})_{L}f(x^{+},x^{3/0},x^{3},x^{-})
\\
&=&-q^2\frac{\lambda }{\lambda _{+}}%
x^{-}D_{q^{2}}^{-}(D_{q^{-2}}^{3}f)(q^{2}x^{3/0},q^{2}q_{+}x^{3})-q^{2}%
\frac{\lambda }{\lambda _{+}}x^{+}D_{q^{2}}^{+}D_{q^{2}q_{-}}^{3}f  \nonumber
\\
&&-{\,}q\frac{\lambda }{\lambda _{+}}%
x^{3/0}D_{q^{2}}^{+}D_{q^{2}}^{-}f(q^{2}q_{-}x^{3})  \nonumber \\
&&-{\,}q^{3}\frac{\lambda ^{2}}{\lambda _{+}}%
x^{+}x^{3/0}D_{q^{2}}^{+}(D_{q^{2}}^{3/0}D_{q^{2}}^{3}f)(q_{+}x^{3}) 
\nonumber \\
&&+\sum_{k=1}^{\infty }\alpha _{+}^{k}\sum_{0\leq i+j\leq k}\Big\{
(M^{+})_{i,j}^{k}(\underline{x})(T_{1}^{0})_{j}^{i}f+q\frac{\lambda }{%
\lambda _{+}}(M^{-})_{i,j}^{k}(\underline{x})(T_{2}^{0})_{j}^{i}f\Big\} 
\nonumber \\
&&-{\,}q^{2}\frac{\lambda }{\lambda _{+}}\sum_{0\leq k+l<\infty }\alpha
_{+}^{k+l}\sum_{i=0}^{k}\sum_{j=0}^{l}\sum_{0\leq u\leq
i+j}(M^{+-})_{i,j,u}^{k,l}(\underline{x})(T_{3}^{0})_{u}^{k,l}f  \nonumber \\
&&+{\,}\frac{\beta }{\lambda _{+}}\sum_{0\leq k+l<\infty }\alpha
_{+}^{k+l+1}\sum_{i=0}^{k+1}\sum_{j=0}^{l}\sum_{0\leq u\leq
i+j}(M^{+-})_{i,j,u}^{k+1,l}(\underline{x})(T_{4}^{0})_{u}^{k,l}f  \nonumber
\\
&&+{\,}\lambda _{+}^{-1}\sum_{0\leq k+l<\infty }\alpha
_{+}^{k+l+1}\sum_{i=0}^{k}\sum_{j=0}^{l+1}\sum_{0\leq u\leq
i+j}(M^{+-})_{i,j,u}^{k,l+1}(\underline{x})(T_{5}^{+})_{u}^{k,l}f,  \nonumber
\end{eqnarray}
where for the purpose of abbreviation we have used the operators \cite{BW01} 
\begin{eqnarray}
\hspace{-0.23in}(T^{3/0})_{j}^{i}f &=&\Big [\left. (O^{3/0})_{i}f\right|
_{x^{3}\rightarrow x^{0}+x^{3/0}}\Big ](q^{2j}x^{3/0}), \\[0.16in]
\hspace{-0.23in}(T_{1}^{-})_{j}^{i}f &=&\Big [\left. (O_{1}^{-})_{i}f\right|
_{x^{3}\rightarrow x^{0}+x^{3/0}}\Big ](q^{2(j+1)}x^{3/0}), \\
\hspace{-0.23in}(T_{2}^{-})_{j}^{i}f &=&\Big [\left. (O_{2}^{-})_{i}f\right|
_{x^{3}\rightarrow x^{0}+x^{3/0}}\Big ](q^{2j}x^{3/0}),  \nonumber \\[0.16in]
\hspace{-0.23in}(T_{1}^{0})_{j}^{i}f &=&\Big [\left. (O_{1}^{0})_{i}f\right|
_{x^{3}\rightarrow y_{+}}\hspace{-0.04in}-q^{2}\frac{\lambda }{\lambda _{+}}%
\left. (O_{2}^{0})_{i}f\right| _{x^{3}\rightarrow q^{2}y_{+}}\Big ]%
(q^{2j}x^{3/0}), \\
\hspace{-0.23in}(T_{2}^{0})_{j}^{i}f &=&\Big [\left. (O_{3}^{0})_{i}f\right|
_{x^{3}\rightarrow x^{0}+x^{3/0}}+\left. (O_{4}^{0})_{i}f\right|
_{x^{3}\rightarrow q^{2}y_{-}}\Big ](q^{2j}x^{3/0}),  \nonumber \\
\hspace{-0.23in}(T_{3}^{0})_{u}^{k,l}f &=&\Big [\left.
(Q_{1}^{0})_{k,l}f\right| _{x^{3}\rightarrow x^{0}+x^{3/0}}\Big ]%
(q^{2u}x^{3/0}),  \nonumber \\
\hspace{-0.23in}(T_{4}^{0})_{u}^{k,l}f &=&\Big [\left.
(Q_{2}^{0})_{k,l}f\right| _{x^{3}\rightarrow x^{0}+x^{3/0}}\Big ]%
(q^{2u}x^{3/0}),  \nonumber \\
\hspace{-0.23in}(T_{5}^{0})_{u}^{k,l}f &=&\Big [\left.
(Q_{3}^{0})_{k,l}f\right| _{x^{3}\rightarrow x^{0}+x^{3/0}}\Big ]%
(q^{2u}x^{3/0}),  \nonumber \\[0.16in]
\hspace{-0.23in}(T_{1}^{+})_{j}^{i}f &=&\Big [\left. (O_{1}^{+})_{i}f\right|
_{x^{3}\rightarrow q^{2}y_{-}}\Big ](q^{2j}x^{3/0}), \\
\hspace{-0.23in}(T_{2}^{+})_{j}^{i}f &=&\Big [\left. (O_{2}^{+})_{i}f\right|
_{x^{3}\rightarrow y_{+}}\Big ](q^{2j}x^{3/0}),  \nonumber \\
\hspace{-0.23in}(T_{3}^{+})_{u}^{k,l}f &=&\Big [\left.
(Q_{1}^{+})_{k,l}f\right| _{x^{3}\rightarrow x^{0}+x^{3/0}}\Big ]%
(q^{2u}x^{3/0}),  \nonumber \\
\hspace{-0.23in}(T_{4}^{+})_{u}^{k,l}f &=&\Big [\left.
(Q_{2}^{+})_{k,l}f\right| _{x^{3}\rightarrow x^{0}+x^{3/0}}\Big ]%
(q^{2u}x^{3/0}).  \nonumber
\end{eqnarray}
The operators $O_{i}^{\mu }$ and $Q_{i}^{\mu }$, $\mu =\pm ,3/0,0,$ as well
as the polynomials $(M^{\pm })_{i,j}^{l}$ and $(M^{+-})_{i,j,u}^{l,m}$ have
already been defined in \cite{BW01}. Their explicit form is once again
listed in appendix \ref{AppA}. In addition, we have set 
\begin{eqnarray*}
\alpha &=&-q^{2}\frac{\lambda ^{2}}{\lambda _{+}^{2}}, \\
\beta &=&q+\lambda _{+}, \\
q_{\pm } &=&1\pm \frac{\lambda }{\lambda _{+}}\frac{x^{3/0}}{x^{3}}, \\
y_{\pm } &=&x^{0}+\frac{2q^{\pm 1}}{\lambda _{+}}x^{3/0}.
\end{eqnarray*}
It is also important to note that the formulae (\ref{DarDerMinAnf}) -(\ref
{DarDerMinEnd}) hold for such functions only which do not explicitly depend
on $x^{0}.$ Due to this fact we have to apply the substitutions 
\begin{equation}
x^{0}\rightarrow x^{3}-x^{3/0}
\end{equation}
to each function the operators in (\ref{DarDerMinAnf}) -(\ref{DarDerMinEnd})
shall act on.

The corresponding formulae for the second differential calculus can in the
usual way be obtained from the above results by the transformations 
\begin{eqnarray}
(\partial ^{\pm })^{-1}\tilde{\triangleright}f &\stackrel{{\QATOP{\pm }{q}}{%
\QATOP{\leftrightarrow }{\leftrightarrow }}{\QATOP{\mp }{1/q}}}{%
\longleftrightarrow }&(\hat{\partial}^{\mp })^{-1}\triangleright f,
\label{TraMinConUnMin} \\
(\partial ^{0})^{-1}\tilde{\triangleright}f &\stackrel{{\QATOP{\pm }{q}}{%
\QATOP{\leftrightarrow }{\leftrightarrow }}{\QATOP{\mp }{1/q}}}{%
\longleftrightarrow }&(\hat{\partial}^{0})^{-1}\triangleright f,  \nonumber \\
(\partial ^{3/0})^{-1}\tilde{\triangleright}f &\stackrel{{\QATOP{\pm }{q}}{%
\QATOP{\leftrightarrow }{\leftrightarrow }}{\QATOP{\mp }{1/q}}}{%
\longleftrightarrow }&(\hat{\partial}^{3/0})^{-1}\triangleright f,  \nonumber
\end{eqnarray}
symbolizing a transition via the substitutions 
\begin{equation}
x^{\pm }\rightarrow x^{\mp },\quad q^{\pm 1}\rightarrow q^{\mp 1},\quad \hat{%
n}^{\pm }\rightarrow -\hat{n}^{\mp }
\end{equation}
\[
D_{q^{a}}^{\pm }\rightarrow D_{q^{-a}}^{\mp },\quad (D_{q^{a}}^{\pm
})^{-1}\rightarrow (D_{q^{-a}}^{\mp })^{-1}. 
\]
Furthermore, one has to realize that the transformations in (\ref
{TraMinConUnMin}) change the normal ordering. This is in complete analogy to
the situation for the representations of the partial derivatives in \cite
{BW01}. The relationship between left and right integrals is now given by 
\begin{eqnarray}
f\triangleleft (\partial ^{0})^{-1}&\stackrel{+\leftrightarrow -}{%
\longleftrightarrow }&-q^{-4}(\hat{\partial}^{0})^{-1}\triangleright f, \\
f\triangleleft (\partial ^{3/0})^{-1}&\stackrel{+\leftrightarrow -}{%
\longleftrightarrow }&-q^{-4}(\hat{\partial}^{3/0})^{-1}\triangleright f, 
\nonumber \\
f\triangleleft (\partial ^{\pm })^{-1}&\stackrel{+\leftrightarrow -}{%
\longleftrightarrow }&-q^{-4}(\hat{\partial}^{\mp })^{-1}\triangleright f, 
\nonumber \\[0.16in]
f\triangleleft (\hat{\partial}^{0})^{-1}&\stackrel{+\leftrightarrow -}{%
\longleftrightarrow }&-q^{4}(\partial ^{0})^{-1}\triangleright f, \\
f\triangleleft (\hat{\partial}^{3/0})^{-1}&\stackrel{+\leftrightarrow -}{%
\longleftrightarrow }&-q^{4}(\partial ^{3/0})^{-1}\triangleright f,  \nonumber
\\
f\triangleleft (\hat{\partial}^{\pm })^{-1}&\stackrel{+\leftrightarrow -}{%
\longleftrightarrow }&-q^{4}(\partial ^{\mp })^{-1}\triangleright f,  \nonumber
\end{eqnarray}
where the symbol $\stackrel{+\leftrightarrow -}{\longleftrightarrow }$ has
the same meaning as in Sect. \ref{DreiEukl}.

Now, we want to deal with the rules for integration by parts which in the
case of left integrals turn out to be 
\begin{eqnarray}
&&\left. (\partial ^{3/0})_{L}^{-1}(\partial _{L}^{3/0}f)\star g\right|
_{x^{3}=0}^{a} \\
&=&f\star g\left\| _{x^{3}=0}^{a}\right. -(\partial ^{3/0})_{L}^{-1}(\Lambda
^{1/2}\tau ^{1}f)\star \left. \partial _{L}^{3/0}g\right| _{x^{3}=0}^{a} 
\nonumber \\
&&+\,q^{1/2}\lambda _{+}^{1/2}\lambda (\partial ^{3/0})_{L}^{-1}(\Lambda
^{1/2}\left( \tau ^{3}\right) ^{-1/2}S^{1}f)\star \left. \partial
_{L}^{+}g\right| _{x^{3}=a}^{b},  \nonumber \\[0.16in]
&&\left. (\partial ^{+})_{L}^{-1}(\partial _{L}^{+}f)\star g\right|
_{x^{-}=0}^{a} \\
&=&f\star g\left\| _{x^{-}=0}^{a}\right. -(\partial ^{+})_{L}^{-1}(\Lambda
^{1/2}\left( \tau ^{3}\right) ^{-1/2}\sigma ^{2}f)\star \left. \partial
_{L}^{+}g\right| _{x^{-}=a}^{b}  \nonumber \\
&&+\,q^{3/2}\lambda _{+}^{-1/2}\lambda (\partial ^{+})_{L}^{-1}(\Lambda
^{1/2}T^{2}f)\star \left. \partial _{L}^{3/0}g\right| _{x^{-}=a}^{b}, 
\nonumber \\[0.16in]
&&\left. (\partial ^{-})_{L}^{-1}(\partial _{L}^{-}f)\star g\right|
_{x^{+}=0}^{a} \\
&=&f\star g\left\| _{x^{+}=0}^{a}\right. -(\partial ^{-})_{L}^{-1}(\Lambda
^{1/2}(\tau ^{3})^{1/2}\tau ^{1}f)\star \left. \partial _{L}^{-}g\right|
_{x^{+}=0}^{a}  \nonumber \\
&&+\,q^{-1/2}\lambda _{+}^{1/2}\lambda (\partial ^{-})_{L}^{-1}(\Lambda
^{1/2}S^{1}f)\star \left. \partial _{L}^{0}g\right| _{x^{+}=0}^{a}  \nonumber
\\
&&+\,\lambda ^{2}(\partial ^{-})_{L}^{-1}(\Lambda ^{1/2}(\tau
^{3})^{-1/2}T^{-}S^{1}f)\star \left. \partial _{L}^{+}g\right| _{x^{+}=0}^{a}
\\
&&-\,q^{-1/2}\lambda _{+}^{-1/2}\lambda (\partial ^{-})_{L}^{-1}(\Lambda
^{1/2}(\tau ^{1}T^{-}-q^{-1}S^{1})f)\star \left. \partial _{L}^{3/0}g\right|
_{x^{+}=0}^{a},  \nonumber \\[0.16in]
&&\left. (\partial ^{0})_{L}^{-1}(\partial _{L}^{0}f)\star g\right|
_{x^{3/0}=0}^{a} \\
&=&f\star g\left\| _{x^{3/0}=0}^{a}\right. -(\partial ^{0})_{L}^{-1}(\Lambda
^{1/2}\sigma ^{2}f)\star \left. \partial _{L}^{0}g\right| _{x^{3/0}=0}^{a} 
\nonumber \\
&&+\,q^{1/2}\lambda _{+}^{-1/2}\lambda (\partial ^{0})_{L}^{-1}(\Lambda
^{1/2}T^{2}(\tau ^{3})^{1/2}f)\star \left. \partial _{L}^{-}g\right|
_{x^{3/0}=0}^{a}  \nonumber \\
&&-\,q^{-1/2}\lambda _{+}^{-1/2}\lambda \left( \partial ^{0}\right)
_{L}^{-1}(\Lambda ^{1/2}(\tau ^{3})^{-1/2}(T^{-}\sigma ^{2}+qS^{1})f)\star
\left. \partial _{L}^{+}g\right| _{x^{3/0}=0}^{a}  \nonumber \\
&&+\,\lambda _{+}^{-1}(\partial ^{0})_{L}^{-1}(\Lambda ^{1/2}(\lambda
_{-}^{2}T^{-}T^{2}+q(\tau ^{1}-\sigma ^{2}))f)\star \left. \partial
_{L}^{3/0}g\right| _{x^{3/0}=0}^{a}.  \nonumber
\end{eqnarray}
In a similar way we get for the conjugated left integrals the identities 
\begin{eqnarray}
&&\left. (\hat{\partial}^{3/0})_{L}^{-1}(\hat{\partial}_{L}^{3/0}f)\star
g\right| _{x^{3}=0}^{a} \\
&=&f\star g\left\| _{x^{3}=0}^{a}\right. -\,(\hat{\partial}%
^{3/0})_{L}^{-1}(\Lambda ^{-1/2}(\tau ^{3})^{-1/2}\sigma ^{2}f)\star \left. 
\hat{\partial}_{L}^{3/0}g\right| _{x^{3}=0}^{a}  \nonumber \\
&&+\,q^{3/2}\lambda _{+}^{1/2}\lambda (\partial ^{3/0})_{L}^{-1}(\Lambda
^{-1/2}T^{2}f)\star \left. \hat{\partial}_{L}^{-}g\right| _{x^{3}=0}^{a}, 
\nonumber \\[0.16in]
&&\left. (\hat{\partial}^{-})_{L}^{-1}(\hat{\partial}_{L}^{-}f)\star
g\right| _{x^{+}=0}^{a} \\
&=&f\star g\left\| _{x^{3}=0}^{a}\right. -(\hat{\partial}^{-})_{L}^{-1}(%
\Lambda ^{-1/2}\tau ^{1}f)\star \left. \hat{\partial}_{L}^{-}g\right|
_{x^{+}=0}^{a}  \nonumber \\
&&+\,q^{1/2}\lambda _{+}^{-1/2}\lambda (\hat{\partial}^{-})_{L}^{-1}(\Lambda
^{-1/2}(\tau ^{3})^{-1/2}S^{1}f)\star \left. \hat{\partial}%
_{L}^{3/0}g\right| _{x^{+}=0}^{a},  \nonumber \\[0.16in]
&&\left. (\hat{\partial}^{+})_{L}^{-1}(\hat{\partial}_{L}^{+}f)\star
g\right| _{x^{-}=0}^{a} \\
&=&f\star g\left\| _{x^{-}=0}^{a}\right. -(\hat{\partial}^{+})_{L}^{-1}(%
\Lambda ^{-1/2}\sigma ^{2}f)\star \left. \hat{\partial}_{L}^{+}g\right|
_{x^{-}=0}^{a}  \nonumber \\
&&+\,q^{1/2}\lambda _{+}^{1/2}\lambda (\hat{\partial}^{+})_{L}^{-1}(\Lambda
^{-1/2}T^{2}(\tau ^{3})^{1/2}f)\star \left. \hat{\partial}_{L}^{0}g\right|
_{x^{-}=0}^{a}  \nonumber \\
&&+\,q^{1/2}\lambda _{+}^{-1/2}\lambda (\hat{\partial}^{+})_{L}^{-1}(\Lambda
^{-1/2}(\tau ^{3})^{-1/2}(T^{+}\sigma ^{2}+q\tau ^{3}T^{2})f)\star \left. 
\hat{\partial}_{L}^{3/0}g\right| _{x^{-}=0}^{a}  \nonumber \\
&&-\,q^{2}\lambda ^{2}(\hat{\partial}^{+})_{L}^{-1}(\Lambda
^{-1/2}T^{2}T^{+}f)\star \left. \hat{\partial}_{L}^{-}g\right|
_{x^{-}=0}^{a},  \nonumber \\[0.16in]
&&\left. (\hat{\partial}^{0})_{L}^{-1}(\hat{\partial}_{L}^{0}f)\star
g\right| _{x^{3/0}=0}^{a} \\
&=&f\star g\left\| _{x^{3/0}=0}^{a}\right. -(\hat{\partial}%
^{0})_{L}^{-1}(\Lambda ^{-1/2}(\tau ^{3})^{1/2}\tau ^{1}f)\star \left. \hat{%
\partial}_{L}^{0}g\right| _{x^{3/0}=0}^{a}  \nonumber \\
&&+\,q^{-1/2}\lambda _{+}^{-1/2}\lambda (\hat{\partial}^{0})_{L}^{-1}(%
\Lambda ^{-1/2}S^{1}f)\star \left. \hat{\partial}_{L}^{+}g\right|
_{x^{3/0}=0}^{a}  \nonumber \\
&&+\,q^{1/2}\lambda _{+}^{-1/2}\lambda (\hat{\partial}^{0})_{L}^{-1}(\Lambda
^{-1/2}(qT^{+}\tau ^{1}-T^{2})f)\star \left. \hat{\partial}_{L}^{-}g\right|
_{x^{3/0}=a}^{b}  \nonumber \\
&&-\,\lambda _{+}^{-1}(\hat{\partial}^{0})_{L}^{-1}(\Lambda ^{-1/2}(\tau
^{3})^{-1/2}(\lambda ^{2}T^{+}S^{1}+q^{-1}(\tau ^{3}\tau ^{1}-\sigma ^{2}))f)
\nonumber \\
&&\star \left. \hat{\partial}_{L}^{3/0}g\right| _{x^{3/0}=a}^{b}.  \nonumber
\end{eqnarray}
In the case of right integrals the rules for integration by parts take the
form 
\begin{eqnarray}
&&(\partial ^{3/0})_{R}^{-1}f\star \left. (\partial _{R}^{3/0}g)\right|
_{x^{3}=0}^{a} \\
&=&f\star g\left\| _{x^{3}=0}^{a}\right. -(\partial ^{3/0})_{R}^{-1}\partial
_{R}^{3/0}f\star \left. (\Lambda ^{-1/2}\sigma ^{2}g)\right| _{x^{3}=0}^{a} 
\nonumber \\
&&-\,q^{1/2}\lambda _{+}^{1/2}\lambda (\partial ^{3/0})_{R}^{-1}(\partial
_{R}^{+}f)\star \left. (\Lambda ^{-1/2}S^{1}g)\right| _{x^{3}=0}^{a}, 
\nonumber \\[0.16in]
&&\left( \partial ^{+}\right) _{R}^{-1}f\star \left. (\partial
_{R}^{+}g)\right| _{x^{-}=a}^{b} \\
&=&f\star g\left\| _{x^{-}=0}^{a}\right. -(\partial ^{+})_{R}^{-1}(\partial
_{R}^{+}f)\star \left. (\Lambda ^{-1/2}(\tau ^{3})^{1/2}\tau ^{1}g)\right|
_{x^{-}=0}^{a}  \nonumber \\
&&-\,q^{3/2}\lambda _{+}^{-1/2}\lambda (\partial ^{+})_{R}^{-1}(\partial
_{R}^{3/0}f)\star \left. (\Lambda ^{-1/2}(\tau ^{3})^{1/2}T^{2}g)\right|
_{x^{-}=0}^{a},  \nonumber \\[0.16in]
&&(\partial ^{-})_{R}^{-1}f\star \left. (\partial _{R}^{-}g)\right|
_{x^{+}=0}^{a} \\
&=&f\star g\left\| _{x^{+}=0}^{a}\right. -(\partial ^{-})_{R}^{-1}(\partial
_{R}^{-}f)\star \left. (\Lambda ^{-1/2}(\tau ^{3})^{-1/2}\sigma
^{2}g)\right| _{x^{+}=0}^{a}  \nonumber \\
&&-\,q^{3/2}\lambda _{+}^{1/2}\lambda (\partial ^{-})_{R}^{-1}(\partial
_{R}^{0}f)\star \left. (\Lambda ^{-1/2}(\tau ^{3})^{-1/2}S^{1}g)\right|
_{x^{+}=0}^{a}  \nonumber \\
&&-\,q^{2}\lambda ^{2}(\partial ^{-})_{R}^{-1}(\partial _{R}^{+}f)\star
\left. (\Lambda ^{-1/2}(\tau ^{3})^{-1/2}T^{-}S^{1}g)\right| _{x^{+}=0}^{a} 
\nonumber \\
&&-\,q^{1/2}\lambda _{+}^{-1/2}\lambda (\partial ^{-})_{R}^{-1}(\partial
_{R}^{3/0}f)\star \left. (\Lambda ^{-1/2}(\tau
^{3})^{-1/2}(S^{1}-qT^{-}\sigma ^{2})g)\right| _{x^{+}=0}^{a},  \nonumber \\%
[0.16in]
&&(\partial ^{0})_{R}^{-1}f\star \left. (\partial _{R}^{0}g)\right|
_{x^{3/0}=0}^{a} \\
&=&f\star g\left\| _{x^{3/0}=0}^{a}\right. -(\partial
^{0})_{R}^{-1}(\partial _{R}^{0}f)\star \left. (\Lambda ^{-1/2}\tau
^{1}g)\right| _{x^{3/0}=0}^{a}  \nonumber \\
&&-\,q^{1/2}\lambda _{+}^{-1/2}\lambda (\partial ^{0})_{R}^{-1}(\partial
_{R}^{-}f)\star \left. (\Lambda ^{-1/2}T^{2}g)\right| _{x^{3/0}=0}^{a} 
\nonumber \\
&&-\,q^{1/2}\lambda _{+}^{-1/2}\lambda (\partial ^{0})_{R}^{-1}(\partial
_{R}^{+}f)\star \left. (\Lambda ^{-1/2}(qS^{1}-\tau ^{1}T^{-})g)\right|
_{x^{3/0}=0}^{a}  \nonumber \\
&&+\,\lambda _{+}^{-1}(\partial ^{0})_{R}^{-1}(\partial _{R}^{3}f)\star
\left. \Lambda ^{-1/2}(\lambda ^{2}T^{2}T^{-}+q(\sigma ^{2}-\tau
^{1}))g\right| _{x^{3/0}=0}^{a}.  \nonumber
\end{eqnarray}
Analogous formulae hold for the conjugated right integrals, as we can write 
\begin{eqnarray}
&&\left. (\hat{\partial}^{3/0})_{R}^{-1}f\star (\hat{\partial}%
_{R}^{3/0}g)\right| _{x^{3}=0}^{a} \\
&=&f\star g\left\| _{x^{3}=0}^{a}\right. -\,(\hat{\partial}^{3/0})_{R}^{-1}(%
\hat{\partial}_{R}^{3/0}g)\star \left. (\Lambda ^{1/2}(\tau ^{3})^{1/2}\tau
^{1}g)\right| _{x^{3}=0}^{a}  \nonumber \\
&&-\,q^{3/2}\lambda _{+}^{1/2}\lambda (\hat{\partial}^{3/0})_{R}^{-1}(\hat{%
\partial}_{R}^{-}f)\star \left. (\Lambda ^{1/2}(\tau
^{3})^{1/2}T^{2}g)\right| _{x^{3}=0}^{a},  \nonumber \\[0.16in]
&&\left. (\hat{\partial}^{-})_{R}^{-1}f\star (\hat{\partial}%
_{R}^{-}g)\right| _{x^{+}=0}^{a} \\
&=&f\star g\left\| _{x^{+}=0}^{a}\right. -(\hat{\partial}^{-})_{R}^{-1}(\hat{%
\partial}_{R}^{-}f)\star \left. (\Lambda ^{1/2}\sigma ^{2}g)\right|
_{x^{+}=0}^{a}  \nonumber \\
&&-\,q^{1/2}\lambda _{+}^{-1/2}\lambda (\hat{\partial}^{-})_{R}^{-1}(\hat{%
\partial}_{R}^{3/0}f)\star \left. (\Lambda ^{1/2}S^{1}g)\right|
_{x^{+}=0}^{a},  \nonumber \\[0.16in]
&&\left. (\hat{\partial}^{+})_{R}^{-1}f\star (\hat{\partial}%
_{R}^{+}g)\right| _{x^{-}=0}^{a} \\
&=&f\star g\left\| _{x^{-}=0}^{a}\right. -(\hat{\partial}^{+})_{R}^{-1}(\hat{%
\partial}_{R}^{+}f)\star \left. (\Lambda ^{1/2}\tau ^{1}g)\right|
_{x^{-}=0}^{a}  \nonumber \\
&&-\,q^{1/2}\lambda _{+}^{1/2}\lambda (\hat{\partial}^{+})_{R}^{-1}(\hat{%
\partial}_{R}^{3/0}f)\star \left. (\Lambda ^{1/2}(\tau ^{1}T^{+}+q^{-1}\tau
^{3}T^{2})g)\right| _{x^{-}=0}^{a}  \nonumber \\
&&-\,\lambda ^{2}(\hat{\partial}^{+})_{R}^{-1}(\hat{\partial}_{R}^{-}f)\star
\left. (\Lambda ^{1/2}T^{2}T^{+}g)\right| _{x^{-}=0}^{a},  \nonumber \\%
[0.16in]
&&\left. (\hat{\partial}^{0})_{R}^{-1}f\star (\hat{\partial}%
_{R}^{0}g)\right| _{x^{3/0}=0}^{a} \\
&=&f\star g\left\| _{x^{3/0}=0}^{a}\right. -(\hat{\partial}^{0})_{R}^{-1}(%
\hat{\partial}_{R}^{0}f)\star \left. (\Lambda ^{1/2}(\tau ^{3})^{1/2}\sigma
^{2}g)\right| _{x^{3/0}=0}^{a}  \nonumber \\
&&-\,q^{3/2}\lambda _{+}^{-1/2}\lambda (\hat{\partial}^{0})_{R}^{-1}(\hat{%
\partial}_{R}^{+}f)\star \left. (\Lambda ^{1/2}(\tau
^{3})^{-1/2}S^{1}g)\right| _{x^{3/0}=0}^{a}  \nonumber \\
&&+\,q^{-1/2}\lambda _{+}^{-1/2}\lambda (\hat{\partial}^{0})_{R}^{-1}(\hat{%
\partial}_{R}^{-}f)\star \left. (\Lambda ^{1/2}(\tau ^{3})^{-1/2}(q\tau
^{3}T^{2}-\sigma ^{2}T^{+})g)\right| _{x^{3/0}=0}^{a}  \nonumber \\
&&-\,q^{-1}\lambda _{+}^{-1}\lambda (\hat{\partial}^{0})_{R}^{-1}(\hat{%
\partial}_{R}^{3/0}f)  \nonumber \\
&&\star \left. (\Lambda ^{1/2}(\tau ^{3})^{1/2}(q^{-1}\lambda
^{2}S^{1}T^{+}+(\tau ^{3})^{-1}\sigma ^{2}-\tau ^{2})g)\right|
_{x^{3/0}=0}^{a}.  \nonumber
\end{eqnarray}

What remains is to deal with the integration over the entire q-Minkowski
space. In this case it is sufficient to consider the classical contributions
in formula (\ref{IntegralM}), as all of the other terms depending on $%
\lambda $ lead to surface terms vanishing at infinity, if we require 
\begin{eqnarray}
\lim_{x^{\mu }\rightarrow \pm \infty }f(\underline{x}) &=&0,\qquad \mu =\pm
,3/0, \\
\lim_{x^{3}\rightarrow \pm \infty }(x^{3})^{n}\left( \frac{\partial }{%
\partial x^{3}}\right) ^{m}f(\underline{x}) &=&0,\qquad \forall \,n,m\in 
\mathbb{N}.  \nonumber
\end{eqnarray}
Thus, we can write 
\begin{eqnarray}
&&(\hat{\partial}^{0})^{-1}(\hat{\partial}^{+})^{-1}(\hat{\partial}%
^{-})^{-1}(\hat{\partial}^{3/0})^{-1}\triangleright f \\
&=&(\hat{\partial}_{(\mu =0)}^{0})^{-1}(\hat{\partial}_{(\mu =0)}^{+})^{-1}(%
\hat{\partial}_{(\mu =0)}^{-})^{-1}(\hat{\partial}_{(\mu
=0)}^{3/0})^{-1}f+S.T.  \nonumber \\
&=&(D_{q^{2}}^{3/0})(D_{q^{2}}^{-})^{-1}(D_{q^{2}}^{+})\left(
(D_{q_{-}q^{2}}^{3})f\right) (q^{-2}x^{+},q^{-2}x^{3})+S.T.  \nonumber
\end{eqnarray}
where S.T stands for neglected surface terms. The above relations follow
from the same reasonings we have already applied to the Euclidean spaces.
Unfortunately, the complexity of our results makes this a rather laborious
task. The different possibilities for composition of the single integrals
are now related to each other by the following identities: 
\begin{eqnarray}
&&(\partial ^{+})^{-1}(\partial ^{3/0})^{-1}(\partial ^{-})^{-1}(\partial
^{0})^{-1}\triangleright f \\
&=&q^{-4}(\partial ^{-})^{-1}(\partial ^{3/0})^{-1}(\partial
^{+})^{-1}(\partial ^{0})^{-1}\triangleright f  \nonumber \\
&=&q^{-2}(\partial ^{3/0})^{-1}(\partial ^{-})^{-1}(\partial
^{+})^{-1}(\partial ^{0})^{-1}\triangleright f  \nonumber \\
&=&q^{-2}(\partial ^{3/0})^{-1}(\partial ^{+})^{-1}(\partial
^{-})^{-1}(\partial ^{0})^{-1}\triangleright f  \nonumber \\
&=&q^{-2}(\partial ^{-})^{-1}(\partial ^{+})^{-1}(\partial
^{3/0})^{-1}(\partial ^{0})^{-1}\triangleright f  \nonumber \\
&=&q^{-2}(\partial ^{+})^{-1}(\partial ^{-})^{-1}(\partial
^{3/0})^{-1}(\partial ^{0})^{-1}\triangleright f.  \nonumber
\end{eqnarray}

As we are used to, integrals over the whole space and their conjugated
versions depend on each other by the transformation rules 
\begin{eqnarray}
&&(\hat{\partial}^{0})^{-1}(\hat{\partial}^{+})^{-1}(\hat{\partial}%
^{-})^{-1}(\hat{\partial}^{3/0})^{-1}\triangleright f \\
&\stackrel{{\QATOP{\pm }{q}}{\QATOP{\leftrightarrow }{\leftrightarrow }}{%
\QATOP{\mp }{1/q}}}{\longleftrightarrow } &(\partial ^{3/0})^{-1}(\partial
^{-})^{-1}(\partial ^{+})^{-1}(\partial ^{0})^{-1}\,\tilde{\triangleright}\,f. 
\nonumber
\end{eqnarray}
If we want to have integrals over the whole space which are built up by
composition of right integrals we can apply the rules 
\begin{eqnarray}
&&f\triangleleft (\partial ^{3/0})^{-1}(\partial ^{-})^{-1}(\partial
^{+})^{-1}(\partial ^{0})^{-1} \\
&\stackrel{+\leftrightarrow -}{\longleftrightarrow } &(\bar{\partial}%
^{0})^{-1}(\bar{\partial}^{+})^{-1}(\bar{\partial}^{-})^{-1}(\bar{\partial}%
^{3/0})^{-1}\triangleright f,  \nonumber \\[0.16in]
&&f\,\tilde{\triangleleft}\,
(\bar{\partial}^{3/0})^{-1}(\bar{\partial}^{-})^{-1}(%
\bar{\partial}^{+})^{-1}(\bar{\partial}^{0})^{-1} \\
&\stackrel{+\leftrightarrow -}{\longleftrightarrow } &(\partial
^{0})^{-1}(\partial ^{+})^{-1}(\partial ^{-})^{-1}(\partial ^{3/0})^{-1}%
\,\tilde{\triangleright}\,f,  \nonumber
\end{eqnarray}
where $(\bar{\partial}^{\mu })^{-1}=-q^{-4}(\hat{\partial}^{\mu })^{-1}.$

Finally, let us come to formulae expressing right and left invariance of our
integrals over the whole space. For this task we shall find it convenient to
introduce the following notation: 
\begin{eqnarray}
\int_{L}d_{q}V{}f &\equiv &(\partial ^{0})^{-1}(\partial ^{+})^{-1}(\partial
^{-})^{-1}(\partial ^{3/0})^{-1}\,\tilde{\triangleright}\,f, \\
\int_{R}d_{q}V{}f &\equiv &f\triangleleft (\partial ^{3/0})^{-1}(\partial
^{-})^{-1}(\partial ^{+})^{-1}(\partial ^{0})^{-1},  \nonumber \\
\int_{L}d_{q}\overline{V}{}f &\equiv &(\bar{\partial}^{0})^{-1}(\bar{\partial%
}^{+})^{-1}(\bar{\partial}^{-})^{-1}(\bar{\partial}^{3/0})^{-1}%
\triangleright f,  \nonumber \\
\int_{R}d_{q}\overline{V}{}f &\equiv &f\,\tilde{\triangleleft}\,
(\bar{\partial}%
^{3/0})^{-1}(\bar{\partial}^{-})^{-1}(\bar{\partial}^{+})^{-1}(\bar{\partial}%
^{0})^{-1}.  \nonumber
\end{eqnarray}
Right and left invariance can then be written as 
\begin{eqnarray}
\partial ^{\mu }\,\tilde{\triangleright}\int_{L/R}d_{q}V{}f
&=&\int_{L/R}d_{q}V{}\partial ^{\mu }\,\tilde{\triangleright}\,f=\varepsilon
(\partial ^{\mu })\int_{L/R}d_{q}V{}f=0, \\
T\triangleright \int_{L/R}d_{q}V{}f &=&\int_{L/R}d_{q}V{}T\triangleright
f=\varepsilon (T)\int_{L/R}d_{q}V{}f=0,  \nonumber \\
K\triangleright \int_{L/R}d_{q}V{}f &=&\int_{L/R}d_{q}V{}K\triangleright
f=\varepsilon (K)\int_{L/R}d_{q}V{}f=\int_{L/R}d_{q}V{}f,  \nonumber \\%
[0.16in]
\left( \int_{L/R}d_{q}V{}f\right) \triangleleft \partial ^{\mu }
&=&\int_{L/R}d_{q}V{}f\triangleleft \partial ^{\mu }=\varepsilon (\partial
^{\mu })\int_{L/R}d_{q}V{}f=0, \\
\left( \int_{L/R}d_{q}V{}f\right) \triangleleft T
&=&\int_{L/R}d_{q}V{}f\triangleleft T=\varepsilon (T)\int_{L/R}d_{q}V{}f=0, 
\nonumber \\
\left( \int_{L/R}d_{q}V{}f\right) \triangleleft K
&=&\int_{L/R}d_{q}V{}f\triangleleft K=\varepsilon
(K)\int_{L/R}d_{q}V{}f=\int_{L/R}d_{q}V{}f,  \nonumber
\end{eqnarray}
where $T$ and $\ K$ denote Lorentz generators from one of the following two
sets: 
\begin{equation}
T\in \{T^{\pm },S^{1},T^{2}\},\qquad K\in \{\tau ^{3},\tau ^{1},\sigma
^{2}\}.
\end{equation}
Anlogous identities hold for the conjugated integrals.

\section{Remarks}

In the past several attempts have been made to introduce the notion of
integration on quantum spaces \cite{Ste96},\cite{CZ93},\cite{Fio93},\cite
{HW92},\cite{Die01}. However, our way of introducing integration is much
more in the spirit of \cite{KM94}, where integration has also been
considered as an inverse of q-differentiation. While in \cite{KM94} the
algebraic point of view dominates, our integration formulae are directly
derived by inversion of q-difference operators. Thus, this method of
integration represents nothing else than a generalization of Jackson's
celebrated q-integral to higher dimensions.

Let us end with a few comments on some typical features of our integrals.
Their expressions are affected by the non-commutative structure of the
underlying quantum spaces in two ways. First of all, integration can always
be reduced to a process of summation. Additionally, there is a correction
term for each order of $\lambda $ vanishing in the undeformed limit $(q=1).$
These new terms result from the finite boundaries of integration and are
responsible for the fact that integral operators referring to different
directions do not commute. However, if we integrate over the whole space,
the single integrals the volume integral is composed of become independent
from each other and can then be expressed by Jackson-Integrals, only.

One important problem which is still open at the moment concerns the
question of integrability. It can easily be seen that for polynomials the
series representing our integrals terminates. Thus, we can say that
polynomials are always integrable. For all other functions which do not vary
too rapidly compared to $\lambda $ it seems to be resonable to assume that
the correction terms become small enough to constitute a convergent series.

\appendix
\section{Notation\label{AppA}}

\begin{enumerate}
\item  The \textit{q-number} is defined by \cite{KS97} 
\begin{equation}
\left[ \left[ c\right] \right] _{q^{a}}\equiv \frac{1-q^{ac}}{1-q^{a}}%
,\qquad a,c\in \mathbb{C}.
\end{equation}
For $m\in \mathbb{N}$, we can introduce the \textit{q-factorial }by setting 
\begin{equation}
\left[ \left[ m\right] \right] _{q^{a}}!\equiv \left[ \left[ 1\right]
\right] _{q^{a}}\left[ \left[ 2\right] \right] _{q^{a}}\ldots \left[ \left[
m\right] \right] _{q^{a}},\qquad \left[ \left[ 0\right] \right]
_{q^{a}}!\equiv 1.
\end{equation}
There is also a q-analogue of the usual binomial coefficients, the so-called 
\textit{q-binomial coefficients} defined by the formula 
\begin{equation}
\QATOPD {\alpha }{m}_{q^{a}}\equiv \frac{\left[ \left[ \alpha \right]
\right] _{q^{a}}\left[ \left[ \alpha -1\right] \right] _{q^{a}}\ldots \left[
\left[ \alpha -m+1\right] \right] _{q^{a}}}{\left[ \left[ m\right] \right]
_{q^{a}}!},
\end{equation}
where $\alpha \in \mathbb{C},$ $m\in \mathbb{N}$.

\item  Note that in functions only such variables are explicitly displayed
as are affected by a scaling. For example, we write 
\begin{equation}
f(q^{2}x^{+})\qquad \text{instead of}\qquad f(q^{2}x^{+},x^{3},x^{-}).
\end{equation}

\item  Arguments enclosed in parentheses refer to the first object on their
left. For example, we have 
\begin{equation}
D_{q^{2}}^{+}f(q^{2}x^{+})=D_{q^{2}}^{+}(f(q^{2}x^{+}))
\end{equation}
or 
\begin{equation}
D_{q^{2}}^{+}(D_{q^{2}}^{+}f+D_{q^{2}}^{-}f)(q^{2}x^{+})=D_{q^{2}}^{+}\left[
(D_{q^{2}}^{+}f+D_{q^{2}}^{-}f)(q^{2}x^{+})\right] .
\end{equation}
However, the symbol $\mid _{x^{\prime }\rightarrow x}$ applies to the whole
expression on its left side reaching up to the next opening bracket or $\pm $
sign.

\item  The \textit{Jackson derivative} referring to the coordinate $x^{A}$
is defined by 
\begin{equation}
D_{q^{a}}^{A}f:=\frac{f(x^{A})-f(q^{a}x^{A})}{(1-q^{a})x^{A}},
\end{equation}
where f may depend on other coordinates as well. Higher Jackson derivatives
are obtained by applying the above operator $D_{q^{a}}^{A}$ several times: 
\begin{equation}
\left( D_{q^{a}}^{A}\right) ^{i}f:=\underbrace{D_{q^{a}}^{A}D_{q^{a}}^{A}%
\ldots D_{q^{a}}^{A}}_{i\text{ times}}f.
\end{equation}

\item  For $a>0,$ $q>1$ and $x^{A}>0,$ the definition of the \textit{Jackson
integral }is 
\begin{eqnarray}
\left. (D_{q^{a}}^{A})^{-1}f\right| _{0}^{x^{A}}
&=&-(1-q^{a})\sum_{k=1}^{\infty }(q^{-ak}x^{A})f(q^{-ak}x^{A}),
\label{Jackson1} \\
\left. (D_{q^{a}}^{A})^{-1}f\right| _{x^{A}}^{\infty }
&=&-(1-q^{a})\sum_{k=0}^{\infty }(q^{ak}x^{A})f(q^{ak}x^{A}), \\
\left. (D_{q^{-a}}^{A})^{-1}f\right| _{0}^{x^{A}}
&=&(1-q^{-a})\sum_{k=0}^{\infty }(q^{-ak}x^{A})f(q^{-ak}x^{A}),  \nonumber \\
\left. (D_{q^{-a}}^{A})^{-1}f\right| _{x^{A}}^{\infty }
&=&(1-q^{-a})\sum_{k=1}^{\infty }(q^{ak}x^{A})f(q^{ak}x^{A}).  \nonumber
\end{eqnarray}
For $a>0,$ $q>1$ and $x^{A}<0,$we set 
\begin{eqnarray}
\left. (D_{q^{a}}^{A})^{-1}f\right| _{x^{A}}^{0}
&=&(1-q^{a})\sum_{k=1}^{\infty }(q^{-ak}x^{A})f(q^{-ak}x^{A}),
\label{Jackson2} \\
\left. (D_{q^{a}}^{A})^{-1}f\right| _{-\infty }^{x^{A}}
&=&(1-q^{a})\sum_{k=0}^{\infty }(q^{ak}x^{A})f(q^{ak}x^{A}),  \nonumber \\
\left. (D_{q^{-a}}^{A})^{-1}f\right| _{x^{A}}^{0}
&=&-(1-q^{-a})\sum_{k=0}^{\infty }(q^{-ak}x^{A})f(q^{-ak}x^{A}),  \nonumber
\\
\left. (D_{q^{-a}}^{A})^{-1}f\right| _{-\infty }^{x^{A}}
&=&-(1-q^{-a})\sum_{k=0}^{\infty }(q^{ak}x^{A})f(q^{ak}x^{A}).  \nonumber
\end{eqnarray}
Note that the formulae (\ref{Jackson1}) and (\ref{Jackson2}) also yield
expressions for q-integrals over any other interval \cite{KS97}.

\item  Additionally, we need operators of the following form 
\begin{equation}
\hat{n}^{A}\equiv x^{A}\frac{\partial }{\partial x^{A}}.
\end{equation}

\item  Calculations for q-deformed Minkowski space show that it is
reasonable to give the following repeatedly appearing polynomials a name of
their own: 
\begin{eqnarray}
(S_{q})_{i,j}(x^{0},x^{3/0}) &\equiv &\left\{ 
\begin{array}{c}
\sum\limits_{p_{1}=0}^{j}\sum\limits_{p_{2}=0}^{p_{1}}\ldots
\sum\limits_{p_{i-j}=0}^{p_{i-j-1}}\prod%
\limits_{l=0}^{i-j}a_{q^{-1}}(x^{0},q^{2p_{l}}x^{3/0}),\quad \\ 
1, \quad \text{if}\; j=i
\end{array}
\right. 
 \nonumber \\
a_{q}\left( x^{0},x^{3/0}\right) &\equiv &-qx^{3/0}(qx^{3/0}+\lambda
_{+}x^{0}), \\
\left( M^{\pm }\right) _{i,j}^{k}(\underline{x}) &\equiv &\left( M^{\pm
}\right) _{i,j}^{k}(x^{0},x^{+},x^{3/0},x^{-})  \nonumber \\
&=&\binom{k}{i}\lambda _{+}^{j}\left( a_{q^{\pm 1}}(q^{2j}x^{3/0})\right)
^{i} \nonumber \\
&&\cdot\,\left( x^{+}x^{-}\right) ^{j}(S_{q})_{k-j,j}(x^{0},x^{3/0}),
  \nonumber
\\
(M^{+-})_{i,j,u}^{k,l}(\underline{x}) &\equiv
&(M^{+-})_{i,j,u}^{k,l}(x^{0},x^{+},x^{3/0},x^{-})  \nonumber \\
&=&\binom{k}{i}\binom{l}{j}\lambda _{+}^{u} (x^{+}x^{-})^{u}\nonumber\\
&&\cdot \left(
a_{q}(q^{2u}x^{3/0})\right) ^{k-i} 
\left( a_{q^{-1}}(q^{2u}x^{3/0})\right)^{l-j} 
\nonumber\\ 
&&\cdot\, (S_{q})_{i+j,u}(x^{0},x^{3/0}). 
\nonumber
\end{eqnarray}

\item  In \cite{BW01} we have introduced the quantities $(K_{\alpha
})_{a_{1},\ldots ,a_{l}}^{(k_{1},\ldots ,k_{l})},$ $k_{i}\in \mathbb{N},$ $%
a_{i},\alpha \in \mathbb{R}$. We also refer to \cite{BW01} for a review of
their explicit calculation. These quantities can be used to define new
operators by setting 
\begin{equation}
D_{a_{1},\ldots ,a_{l}}^{\left( k_{1},\ldots ,k_{l}\right) }x^{n}=\left\{ 
\begin{array}{c}
(K_{n})_{a_{1},\ldots ,a_{l}}^{\left( k_{1},\ldots ,k_{l}\right)
}x^{n-k_{1}-\ldots -k_{l}}, \\ 
0,\qquad \text{if }n<k_{1}+\ldots +k_{l},
\end{array}
\right.
\end{equation}
Especially, in the case of q-deformed Minkowski space we need the operators 
\begin{eqnarray}
(D_{1,q}^{3})^{k,l} &\equiv &D_{1,q^{2}}^{(k,l)}, \\
(D_{2,q}^{3})^{k,l} &\equiv &D_{y_{-}/x^{3},q^{2}y_{-}/x^{3}}^{(k,l)}, 
\nonumber \\
(D_{3,q}^{3})_{i,j}^{k,l} &\equiv
&D_{y_{+}/x^{3},q^{2}y_{+}/x^{3},y_{-}/x^{3},q^{2}y_{-}/x^{3}}^{(k,l,i,j)}, 
\nonumber
\end{eqnarray}
where 
\begin{equation}
y_{\pm }=y_{\pm }(x^{0},x^{3/0})=x^{0}+\frac{2q^{\pm 1}}{\lambda _{+}}x^{3/0}.
\end{equation}
Notice that these operators have to act upon the coordinate $x^{3}$, only.

\item  In Sect. \ref{KapMin} the representations of $(\hat{\partial}^{\mu
})^{-1},$ $\mu =\pm ,0,3/0,$ have been formulated by using the operators 
\begin{eqnarray}
(O^{3/0})_{k}f &=&(D_{2,q}^{3})^{k,k+1}f(q^{2}x^{+}), \\[0.16in]
(O_{1}^{-})_{k}f &=&x^{-}(D_{2,q}^{3})^{k+1,k+1}f(q^{2}x^{+}), \\
(O_{2}^{-})_{k}f &=&x^{3/0}D_{q^{2}}^{+}(D_{2,q}^{3})^{k,k+1}f(q^{2}x^{+}), 
\nonumber \\[0.16in]
(O_{1}^{0})_{k}f &=&D_{q^{2}}^{3/0}(D_{1,q}^{3})^{k,k}f \\
&&-\,q^{3}\lambda _{+}^{-1}\lambda
^{2}x^{+}x^{3/0}D_{q^{2}}^{+}D_{q^{2}}^{3/0}(D_{1,q}^{3})^{k,k+1}f, 
\nonumber \\
(O_{2}^{0})_{k}f
&=&x^{-}(D_{1,q^{-1}}^{3})^{k,k+1}D_{q^{2}}^{-}f(q^{2}x^{3/0}),  \nonumber \\
(O_{3}^{0})_{k}f &=&qx^{+}D_{q^{2}}^{+}(D_{2,q}^{3})^{k,k+1}f,  \nonumber \\
(O_{4}^{0})_{k}f
&=&x^{3/0}D_{q^{2}}^{+}(D_{1,q^{-1}}^{3})^{k,k}D_{q^{2}}^{-}f,  \nonumber \\%
[0.16in]
(Q_{1}^{0})_{k,l}f &=&(x^{0}+q^{-1}\lambda
x^{3})(D_{3,q}^{3})_{l,l+1}^{k+1,k}f \\
&&+\,q\lambda _{+}^{-1}\lambda (q+\lambda
_{+})x^{+}x^{3/0}D_{q^{2}}^{+}(D_{3,q}^{3})_{l,l+1}^{k,k+1}f  \nonumber \\
&&-\,q^{3}\lambda _{+}^{-1}\lambda ^{2}x^{+}x^{3/0}(x^{0}+q^{-1}\lambda
x^{3/0})D_{q^{2}}^{+}(D_{3,q}^{3})_{l,l+1}^{k+1,k+1}f,  \nonumber \\
(Q_{2}^{0})_{k,l}f &=&(D_{3,q}^{3})_{l,l+1}^{k+1,k+1}f,  \nonumber \\
(Q_{3}^{0})_{k,l}f &=&q^{-1}(D_{3,q}^{3})_{l+1,l+1}^{k+1,k}f-q^{2}\lambda
_{+}^{-1}\lambda
^{2}x^{+}x^{3/0}D_{q^{2}}^{+}(D_{3,q}^{3})_{l+1,l+1}^{k+1,k+1}f,  \nonumber
\\[0.16in]
(O_{1}^{+})_{k}f &=&(D_{1,q^{-1}}^{3})^{k,k}D_{q^{2}}^{-}f, \\
(O_{2}^{+})_{k}f &=&x^{+}D_{q^{2}}^{3/0}(D_{1,q}^{3})^{k,k+1}f,  \nonumber \\%
[0.16in]
(Q_{1}^{+})_{k,l}f &=&(q+\lambda _{+})x^{+}(D_{3,q}^{3})_{l,l+1}^{k,k+1}f \\
&&-\,q^{2}\lambda x^{+}(x^{0}+q^{-1}\lambda
x^{3})(D_{3,q}^{3})_{l,l+1}^{k+1,k+1}f,  \nonumber \\
(Q_{2}^{+})_{k,l}f &=&x^{+}(D_{3,q}^{3})_{l+1,l+1}^{k+1,k+1}f.  \nonumber
\end{eqnarray}
\end{enumerate}

\noindent \textbf{Acknowledgement}\newline
First of all I want to express my gratitude to Julius Wess for his efforts,
suggestions and discussions. Also I would like to thank Michael Wohlgenannt,
Fabian Bachmeier, Christian Blohmann, and Marcus Dietz for useful
discussions and their steady support.


\begin{thebibliography}{99}
\bibitem{RTF90}  N.Yu. Reshetikhin, L.A. Takhtadzhyan, L.D. Faddeev, \textit{%
Quantization of Lie Groups and Lie Algebras}, Leningrad Math. J. \textbf{1}
(1990) 193.

\bibitem{Wes00}  J. Wess, \textit{q-deformed Heisenberg Algebras}, in H.
Gausterer, H. Grosse and L.Pittner, eds., Proceedings of the 38.
Internationale Universit\"{a}tswochen f\"{u}r Kern- und Teilchen physik, no.
543 in Lect. Notes in Phys., Springer -Verlag, Schladming (2000), math-phy
9910013.

\bibitem{Maj93}  S. Majid,\textit{\ Braided geometry: A new approach to
q-deformations, }in R. Coquereaux et al., eds., First Caribbean Spring
School of Mathematics and Theoretical\ Physics, World Sci., Guadeloupe
(1993).

\bibitem{Moy49}  J.E. Moyal, \textit{Quantum mechanics as a statistical
theory, }Proc. Camb. Phil. Soc. \textbf{45} (1949) 99.

\bibitem{MSSW00}  J. Madore, S. Schraml, P. Schupp, J. Wess, \textit{Gauge
Theory on Noncommutative Spaces, }Eur. Phys. J. C\textbf{16} (2000) 161,
hep-th/0103120.

\bibitem{WW01}  H. Wachter, M. Wohlgenannt, \textit{*-Products on Quantum
Spaces},\textit{\ }Eur. Phys. J. \textbf{C23 }(2002) 761-767, hep-th/0103120.

\bibitem{Dri85}  V.G. Drinfeld, \textit{Hopf algebras and the quantum
Yang-Baxter equation, }Soviet Math. Dokl. \textbf{32} (1985) 254-258.

\bibitem{Jim85}  M. Jimbo, \textit{A q-analogue of U(g) and the Yang-Baxter
equation,} Lett. Math. Phys. \textbf{10} (1985) 63-69.

\bibitem{WZ91}  J. Wess, B. Zumino,\textit{\ Covariant differential calculus
on the quantum hyperplane}, Nucl. Phys. B. Suppl. \textbf{18} (1991),
302-312.

\bibitem{BW01}  C. Bauer, H. Wachter, \textit{Operator representations on
quantum spaces, }, Eur. Phys. J. C \textbf{31} (2003) 261, math-ph/0201023.

\bibitem{Jac27}  F.H. Jackson, \textit{q-Integration,} Proc. Durham Phil.
Soc. \textbf{7} (1927), 182-189.

\bibitem{LWW97}  A. Lorek, W. Weich, J. Wess, \textit{Non Commutative
Euclidean and Minkowski Structures, }Z. Phys. C\textbf{76} (1997) 375,
q-alg/9702025.

\bibitem{KS97}  A. Klimyk, K. Schm\"{u}dgen, \textit{Quantum Groups and
their Representations,} Springer Verlag, Berlin (1997).

\bibitem{Maj95}  S. Majid, \textit{Foundations of Quantum Group Theory, }%
University Press, Cambridge (1995).

\bibitem{Oca96}  H. Ocambo, \textit{SO}$_{q}$(4) \textit{quantum mechanics},
Z. Phys. C\textbf{70} (1996) 525.

\bibitem{CSSW90}  U. Carow-Watamura, M. Schlieker, M. Scholl, S. Watamura,
Z. Phys. \textbf{C48} (1990) 159.

\bibitem{SWZ91}  W.B. Schmidtke, J. Wess, B. Zumino, \textit{A q-deformed
Lorentz Algebra in Minkowski phase space, }Z. Phys. C\textbf{52 }(1991) 471.

\bibitem{Maj91}  S. Majid, \textit{Examples of braided groups and braided
matrices,} J. Math. Phys. \textbf{32} (1991) 3246-3253.

\bibitem{Dob94}  V. K. Dobrev, \textit{New q-Minkowski space-time and
q-Maxwell equations hierarchy from q-conformal invariance,} Phys. Lett. 
\textbf{341B} (1994) 133-138 \& \textbf{346B} (1995) 427.

\bibitem{OSWZ92}  O. Ogievetsky, W.B. Schmittke, J. Wess, B. Zumino, \textit{%
q-deformed Poincar\'{e} Algebra,} Commun. Math. Phys. \textbf{150} (1992)
495.

\bibitem{RW99}  M. Rohregger, J. Wess, \textit{q-Deformed Lorentz Algebra in
Minkowski phase space, }Eur. Phys. J.\textbf{7} (1999) 177.

\bibitem{Ste96}  H. Steinacker, \textit{Integration on quantum Euclidean
Space and sphere,} J. Math. Phys. \textbf{37} (1996) 4738.

\bibitem{CZ93}  C. Chryssomalakos, B. Zumino, \textit{Translations integrals
and Fourier transforms in the quantum plane, }preprint LBL-34803,
UCCB-PTH-93/30, in Salamfestschrift, edited by A. Ali, J. Ellis, and S.
Randjbar-Daemi, World Sci., Singapore, 1993.

\bibitem{Fio93}  G. Fiore, \textit{The SO}$_{q}$\textit{(N)-symmmetric
harmonic oszillator on the quantum Euclidean space...,} Int. J. Mod. Phys., A%
\textbf{8} (1993) 4679.

\bibitem{HW92}  A. Hebecker, W. Weich, \textit{Free particle in q-deformed
configuration space, }Lett. Math. Phys. \textbf{26} (1992) 245.

\bibitem{Die01}  M.A. Dietz, \textit{Symmetrische Formen auf
Quantenalgebren, }diploma-theses, Universit\"{a}t Hamburg, Fakult\"{a}t
f\"{u}r Physik (2001).

\bibitem{KM94}  A. Kempf, S. Majid, \textit{Algebraic q-integration and
Fourier theory on quantum and braided spaces, }J. Math. Phys. \textbf{35}
(1994) 6802.
\end{thebibliography}
\end{document}